\documentclass[twocolumn,aps,prb,amsmath,amssymb,superscriptaddress,longbibliography]{revtex4-2}

\usepackage[backref=none,
bookmarksnumbered=true,
bookmarks=true,
bookmarksopen=true,
colorlinks=true,
citecolor=blue,
linkcolor=blue,
anchorcolor=green,
urlcolor=blue,unicode=false]{hyperref}
\usepackage[table]{xcolor}
\usepackage{graphicx}
\usepackage{epstopdf}
\usepackage{physics}
\usepackage{bbold}
\usepackage{amsmath}

\begin{document}
\title{Floquet product mode and eigenphase order}

\author{Felix Möckel}
\affiliation{\mbox{Dahlem Center for Complex Quantum Systems and Fachbereich Physik, Freie Universit\"at Berlin, 14195 Berlin, Germany}}
\author{Harald Schmid}
\affiliation{\mbox{Dahlem Center for Complex Quantum Systems and Fachbereich Physik, Freie Universit\"at Berlin, 14195 Berlin, Germany}}
\affiliation{\mbox{Technical University of Munich, TUM School of Natural Sciences, Physics Department, 85748 Garching, Germany}}
\affiliation{Munich Center for Quantum Science and Technology (MCQST), Schellingstr.\ 4, 80799 M{\"u}nchen, Germany}
\author{Felix von Oppen}
\affiliation{\mbox{Dahlem Center for Complex Quantum Systems, Fachbereich Physik, and Halle-Berlin-Regensburg}\\ Cluster of Excellence CCE, Freie Universit\"at Berlin, 14195 Berlin, Germany}

\begin{abstract}
    We study the robustness of the Floquet quantum Ising model against integrability-breaking perturbations, focusing on the phase hosting both Majorana zero and $\pi$ modes. A recent work [Phys. Rev. B \textbf{110}, 075117, (2024)] observed that the Floquet product mode, a composite edge mode constructed from both Majorana operators, is considerably more robust than the individual Majorana edge modes. We analyze these strong modes from the point of view of the eigenphase order present in finite chains with open boundary conditions. As a result of the Majorana modes, all Floquet eigenstates come in quadruplets in the integrable limit. We show that the robustness of the various modes as well as the behavior of the boundary spin correlation functions can be understood in terms of the spectral statistics of these quadruplets in the presence of integrability-breaking perturbations. 
\end{abstract} 

\maketitle
\section{Introduction}

There has recently been much interest in strong modes in quantum many-body systems \cite{Fendley2012,Fendley2014,Fendley2016,Vasiloiu2018,Vasiloiu2019,Kemp2020,Kemp2017,Chepiga2023, Laflorencie2023,Essler2025, Moessner2016,Keyserlingk2016, Yates2020,Yates2019,Parker2019,Yates2020,Yates2021,Mi2022,Yeh2023,Yeh2024,Vernier2024,Schmid2024,Jin2025,Friedman2022,Schmid2024a,Klobas2023}. Strong zero modes are edge operators which commute with the Hamiltonian up to exponentially small corrections in the system size \cite{Fendley2012,Fendley2016,Kemp2017,Vasiloiu2018,Vasiloiu2019,Kemp2020,Laflorencie2023,Chepiga2023,Essler2025}. A prominent example are the Majorana zero modes in the Kitaev chain and their cousins in the quantum Ising model \cite{Kitaev2001,Jiang2011,Dutta2013,Bauer2019}. These modes are referred to as strong in view of their robustness against integrability-breaking perturbations, provided that protecting symmetries are preserved. In Floquet systems, strong zero modes are complemented by strong $\pi$ modes  \cite{Moessner2016,Yates2020,Yates2019,Parker2019,Yates2020,Yates2021,Mi2022,Yeh2023,Yeh2024,Vernier2024,Schmid2024,Jin2025}. While strong zero modes commute with the Floquet operator, strong $\pi$ modes anticommute. A paradigmatic model hosting both strong zero and $\pi$ modes is the Floquet quantum Ising model. 

\begin{figure*}[t!]
    \centering
    \includegraphics[width=\linewidth]{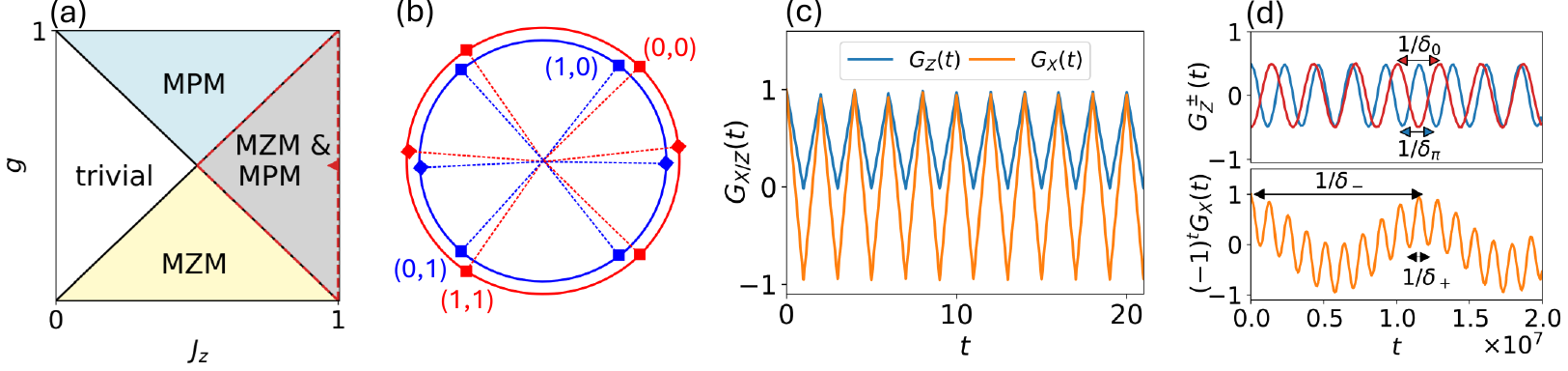}
    \caption{Floquet dynamics in the integrable case for the phase exhibiting both Majorana zero and $\pi$ modes. (a) Phase diagram of the Floquet quantum Ising chain, with phases exhibiting Majorana zero modes (MZM) and Majoranas $\pi$ modes (MPM). We focus on the $0\pi$ phase where both modes are present (red lines). (b) Eigenphase order of the Floquet many-body spectrum in the $0\pi$ phase: Eigenphases form quadruplets, labeled by their boundary mode occupation $(n_{0},n_\pi)$ in the different parity sectors (colors). (c) Correlation functions $G_{X,Z}(t)$ of the boundary operators $X_1$ and $Z_1$ show period doubling.  (d) Coherent long-time oscillations of $G^\pm_Z(t)$ (defined in the text) and  $(-1)^tG_X(t)$ are induced by hybridization of boundary modes, with oscillation period determined by inverse Majorana-hybridization splittings [Fig.\ \ref{figsplit}(a)]. Parameters: $N=8$, $J_z=0.9$, $g=0.505$.
    }
    \label{fig1}
\end{figure*}

The phase diagram of the Floquet quantum Ising model is shown in Fig.\ \ref{fig1}(a). The phases labeled MZM and MPM host Majorana zero and $\pi$ modes, respectively, at the ends of a finite chain, protected by the Ising spin flip symmetry. The phases can be characterized by the eigenphase spectrum of the Floquet operator and differ in their eigenphase order [Fig.\ \ref{fig1}(b)]. The Majorana zero modes (MZMs) induce pairs of eigenstates with degenerate eigenphases, while Majorana $\pi$ modes (MPMs) result in antipodal pairs, whose eigenphases differ by $\pi$. This pairing persists throughout the entire many-body spectrum and holds up to corrections, which are exponentially small in the system size. Much work has been expended in investigating the stability of the Majorana modes  in the presence of integrability-breaking perturbations, both theoretically and experimentally \cite{Kemp2017,Kemp2020,Chepiga2023,Yates2020,Yates2019,Parker2019,Schmid2024,Jin2025,Vernier2024}. Here, we focus on the phase exhibiting both MZMs and MPMs ($0\pi$ phase). 

This problem has recently been brought into focus in a paper by Yeh, Rosch, and Mitra \cite{Yeh2024}. Unlike the phases hosting only MZMs or MPMs, the $0\pi$ phase is a paramagnetic phase, so that the eigenphase order is not accompanied by eigenstate order \cite{Huse2013,Pekker2014,Khemani2016,mi2022time}. Nevertheless, Ref.\ \cite{Yeh2024} establishes that there is interesting strong mode physics in this phase. The product of the Majorana operators constitutes an independent edge mode in the presence of integrability-breaking perturbations. This Floquet product mode can be probed directly through local boundary correlation functions. Remarkably, the product mode persists up to much larger integrability-breaking perturbations than the individual MZMs and MPMs. Reference \cite{Yeh2024} explains this in terms of Fermi's golden rule. 

The purpose of this paper is to interpret the product mode from the point of view of eigenphase order. The spectral pairings imply that in the $0\pi$ phase, the many-body eigenstates of the Floquet quantum Ising model form spectral quadruplets. We investigate these spectral quadruplets in the presence of an integrability-breaking perturbation and show that the level statistics of splittings away from perfect zero and $\pi$ pairings is directly related to the temporal boundary spin correlations. This can then be used to characterize the stability of the Majorana modes as well as the Floquet product mode, when integrability is broken. Our approach allows one to gain insight into the full form of the boundary correlation functions beyond their decay rate. In particular, we find that the decay of the correlation function characterizing the Floquet product mode is Gaussian rather than exponential in the limit of weak breaking of  integrability. 

The remainder of this paper is organized as follows. Section \ref{sec:model} reviews the phase diagram and the eigenphase order of the quantum Ising model. Section \ref{sec:BreakIntegrability} considers the eigenphase order in the presence of an integrability-breaking interaction and contains the central results of this work. Numerical results are presented in Sec.\ \ref{sec:numerical}. Analytical considerations based on perturbation theory in the strength of the integrability-breaking term are given in Sec.\ \ref{sec:analytical}.
We conclude in Sec.\ \ref{sec:conclusions}.

\section{Model and integrable limit} 
\label{sec:model}

We study the Floquet spin chain 
\cite{Khemani2016,Kemp2017,Yates2019}, 
\begin{align}
     U_F =
     &\exp\bigg\{\frac{i\pi J_z}{2} \sum^{N-1}_{j=1} Z_jZ_{j+1} \bigg\}    
     \notag
     \\
      \times & \exp\bigg\{\frac{i \pi g}{2} \sum^N_{j=1} X_j 
     \bigg\} \exp\bigg\{\frac{i\pi J_x}{2} \sum^{N-1}_{j=1} X_jX_{j+1} \bigg\}
     \label{eq:UF}
\end{align} 
with open boundary conditions. Here, $X_j$ and $Z_j$ denote Pauli matrices at site $j$, $J_{x}$ and $J_z$ are the strengths of the exchange couplings in the $x$- and $z$-directions, and $g$ denotes the transverse field. The Floquet operator $U_F$ governs the stroboscopic time evolution $\ket{\Psi(t)}=(U_F)^t\ket{\Psi(0)}$. Due  to the integer nature of time $t$, the Floquet many-body spectrum defined by $U_F\ket{n}=e^{-iE_n}\ket{n}$ can be restricted to the range $E_n \in [-\pi,\pi]$. The Floquet operator commutes with the global spin-flip operator $P=\prod^{N}_{j=1} X_j$. 

For $J_x=0$, the model reduces to the Floquet quantum Ising model, $U_{F,0}=U_F(J_x=0)$. In this limit, the model is integrable and the Floquet operator can be diagonalized by means of the Jordan-Wigner transformation 
\begin{equation}
  X_j=e^{i\pi c^\dagger_jc_j^{\phantom{\dagger}}} \quad ; \quad    Z_j = e^{i\pi \sum_{l<j}c^\dagger_{l} c_{l}^{\phantom{\dagger}}}(c^\dagger_j+c_j^{\phantom{\dagger}})
\end{equation}
to fermionic operators $c_j$ \cite{Dutta2013,Khemani2016,Yates2019}. For an infinite chain, the excitations of the resulting free-fermion model (Floquet-Kitaev chain) are momentum eigenstates with  single-particle spectrum
\begin{align}
    \cos \epsilon_k = \cos(\pi g) \cos(\pi J_z)+ \sin(\pi g) \sin(\pi J_z) \cos k,
    \label{eq:excitation spectrum}
\end{align}
where $k \in [-\pi,\pi]$. The spectrum is invariant under $g\to g\pm 2$, $J_z\to J_z\pm 2$ as well as $J_z\to -J_z$ and the combined operation $g\to -g$ and $k\to -k$. One identifies topological phase transitions with associated gap closings at $k=0$ and $k=\pi$, when $g=J_z$ and $g=1-J_z$ (as well as symmetry-related lines), respectively. These phase boundaries and their symmetry-induced partners underlie the phase diagram of the Floquet quantum Ising model shown in Fig.\ \ref{fig1}(a).

In long but finite chains with open boundary conditions, the four topologically distinct phases of the fermionized model are signaled by the absence or presence of two types of Majorana end modes as indicated in the phase diagram. MZMs appear across the phase boundaries parallel to $g=J_z$, while MPMs appear at phase boundaries parallel to $g=1-J_z$. In this paper, we focus on the phase with both MZMs and MPMs, which is marked by a red line in Fig.\ \ref{fig1}(a). (Notice that the phase extends symmetrically about the line $J_z=1$.) Up to finite-size splittings, the spectrum in this phase is composed of quadruplets $\ket{(n_0,n_\pi),n}$ with eigenphases $E_{(n_0,n_\pi),n}$, where the
$n_{0}, n_{\pi}\in \{0,1\}$ distinguish states within the quadruplets and $n$ enumerates the quadruplets [Fig.\ \ref{fig1}(b)]. The existence of MZMs implies that all many-body eigenphases come in degenerate pairs. Moreover, the existence of MPMs implies that all eigenphases come in pairs, which differ by $\pi$. We can thus choose $n_0$ and $n_\pi$ as the occupations of the fermion modes composed from the two MZMs and MPMs, respectively.

\begin{figure}[b!]
    \centering
    \includegraphics[width=\linewidth]{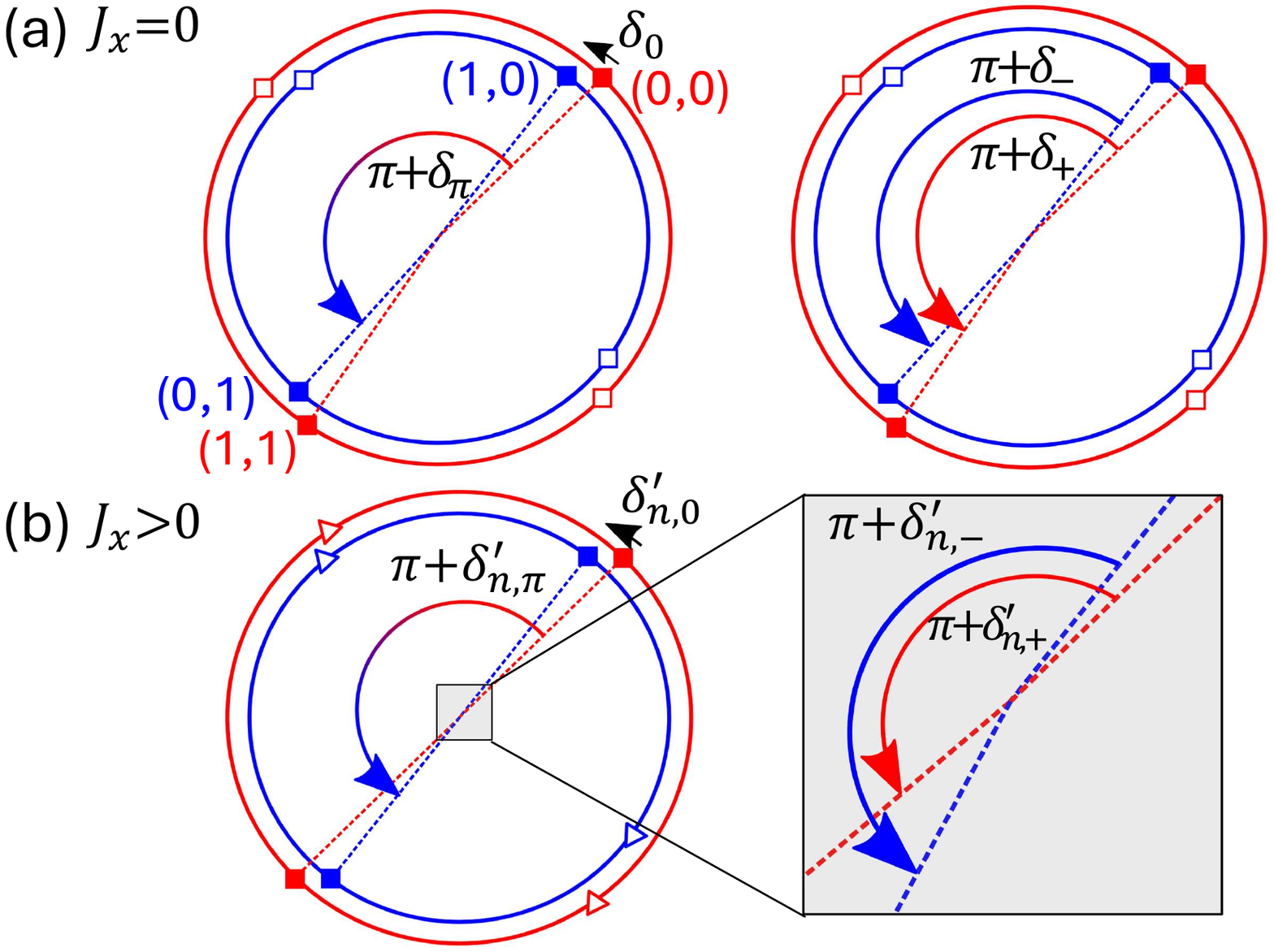}
    \caption{Quadruplet splittings in the many-body Floquet spectrum (not to scale). (a) Non-interacting case. Quadruplets are labeled by their boundary mode occupation $(n_{0},n_\pi)$ in the different parity sectors (see colors).  Spectral pairings are hybridization-split described through inter-parity splittings $\delta_{0,\pi}$ (left) or  intra-parity splittings $\delta_\pm$ (right). Inter- and intra-parity splittings are comparable in magnitude and splittings are identical for all quadruplets. (b) Interacting case. Splittings can be distinguished spectrally between parity sectors and depend on the quadruplet $n$. Inter-parity splittings (left) are much larger than intra-parity splittings (right zoom).}
    \label{figsplit}
\end{figure}

The absence or presence of boundary modes is heralded by the boundary spin correlation functions 
\begin{align}
    G_O(t)=\expval{O(t)O(0)}=\frac{1}{2^N}\sum_{n m}|O_{n m}|^2
    e^{-i(E_n-E_m)t} 
    \label{eq:corrfunc}
\end{align}
for $ O=X_1,Z_1 $. Here, we define $O(t) = (U^\dagger_F)^t O U_F^t$ and $\expval{\dots}=2^{-N}\mathrm{tr}[\dots] $. The correlation function probes the entire many-body Hilbert space uniformly akin to an infinite-temperature ensemble. The quantum dynamics of the phase with both MZMs and MPMs becomes particularly transparent at the sweet spot $g=0.5$ and $J_z=1$ [red triangle in Fig.\ \ref{fig1}(a)], where the Majorana modes are perfectly localized. At the leftmost site of the chain, the Jordan-Wigner transformation does not involve a string operator and we can identify the Majorana operators as $\gamma_{0,L}=(Z_1+Y_1)/\sqrt{2}$ (MZM) and $\gamma_{\pi,L}=(Z_1-Y_1)/\sqrt{2}$ (MPM). We observe that $Z_1$ measures an equal-weight linear combination of the two Majorana modes. The MZM (MPM) commute (anticommute) with the Floquet operator, so that 
\begin{align}    U_F^\dagger\gamma_{0,L}U_F=\gamma_{0,L}, \qquad U_F^\dagger\gamma_{\pi,L}U_F=-\gamma_{\pi,L}.
    \label{eq:comm 0 pi mode}
\end{align}
Hence, the MZM contributes a time-independent constant to the correlation function, while the contribution of the MPM alternates in sign at subsequent times. Correspondingly, one observes that $G_Z(t)$ alternates between $0$ and $1$ [Fig.\ \ref{fig1}(c)]. In contrast, $X_1=i \gamma_{0,L} \gamma_{\pi,L}$  measures the product of both Majorana modes, $\chi_{L} = i \gamma_{0,L}\gamma_{\pi,L}$, which anticommutes with the Floquet operator and therefore satisfies
\begin{align}
    U_F^\dagger\chi_{L}U_F=-\chi_{L}.
    \label{eq:comm prod mode}
\end{align}
As a result, one finds that the correlation function $G_X(t)$ alternates between $\pm 1$ [Fig.\ \ref{fig1}(c)]. 

\begin{figure*}[t!]
    \centering
    \includegraphics[width=\linewidth]{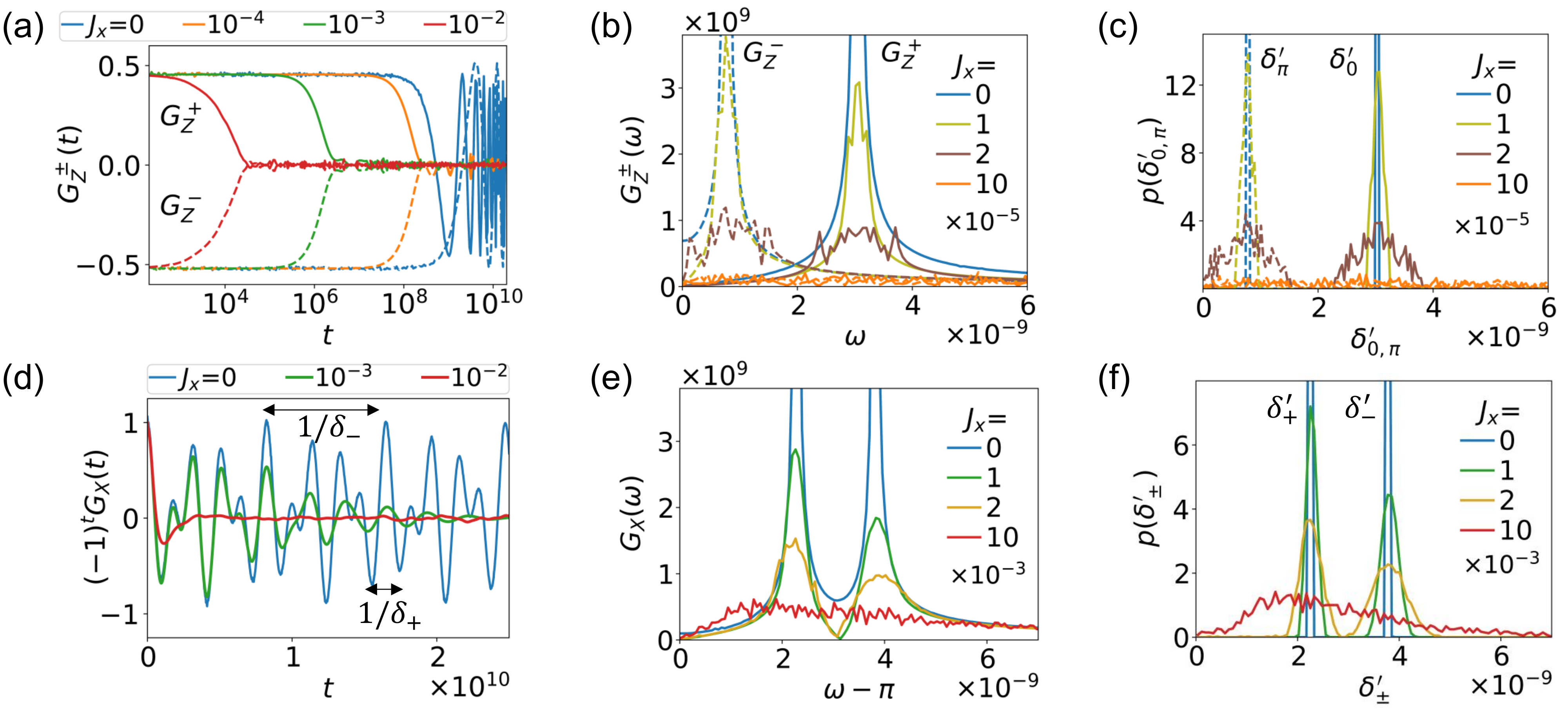}
    \caption{
    Floquet quadruplet dynamics (interacting case). (a)  Correlation functions $G^+_Z(t)$ (solid) and $G^-_Z(t)$ (dashed). The dynamics relaxes quickly for small interactions. (b) Fourier transforms $G^+_Z(\omega)$ (solid) and $G^-_Z(\omega)$ (dashed). Peaks associated to the intra-parity splittings $\delta_{0,\pi}$ vanish with increasing coupling. (c) Inter-parity splitting distributions for zero (solid) and $\pi$-splittings (dashed). The behavior tracks $G^{\pm}_Z(\omega)$. 
    (d) Correlation function $G_X(t)$. Long-time oscillations are visible for much larger interactions than for $G_Z(t)$. (e) Fourier transform $G_X(\omega)$. Peaks associated to the inter-parity splittings $\delta_{\pm}$ can still be identified in the presence of moderate interactions. (f) Intra-parity splittings.  The  behavior follows $G_X(\omega)$. Parameters: $N=12$, $J_z=0.9$, $g=0.52$.}
    \label{fig3}
\end{figure*}

Away from the sweet spot, the Majorana modes have a finite spatial extent and develop a hybridization splitting away from zero (MZM) or $\pi$ (MPM), which is exponentially small in the chain length and uniform across all quadruplets $n$. The large overlap of the Majorana modes with the boundary spins implies that the behaviors of the boundary correlation functions persist qualitatively at short times. At exponentially long times, the period-two oscillations are superimposed by slow oscillations with periods given by the inverse hybridization splittings, see Fig.\ \ref{fig1}(d). Moreover, the (anti-)commutation relations underlying Eqs.\ \eqref{eq:comm 0 pi mode} and \eqref{eq:comm prod mode} remain valid up to exponential accuracy in the chain length in the vicinity of the sweet spot.

We find it useful to define the staggered combinations 
$G^\pm_Z(t)=\frac{1}{2}(G_Z(2t+1)\pm G_Z(2t))$, which disentangle the oscillation periods associated with the hybridization splittings of the MZMs and MPMs and eliminate the period-two oscillations [see upper panel of Fig.\ \ref{fig1}(d)]. The correlation function $G^+_Z(t)$ emphasizes the time-independent contribution of the MZMs and consequently exhibits long-time oscillations with period $2\pi/\delta_{0}$, where $\delta_0 \propto e^{-N/\xi_{0}}$ is the hybridization splitting of the MZMs. The correlation function $G^-_Z(t)$ emphasizes the amplitude of the period-two oscillations and exhibits long-time oscillations with period  $2\pi/\delta_{\pi}$, where $\delta_{\pi} \propto e^{-N/\xi_{\pi}}$ is the hybridization splitting of the MPMs. Here, $\xi_{0}$ and $\xi_{\pi}$ denote the Majorana localization lengths \cite{Lerose2021,Schmid2024}
\begin{align}
 \xi_{0,\pi}=-\frac{1}{\ln \lambda_{0,\pi} },
 \quad \lambda_0=\frac{\tan(\frac{\pi g}{2})}{\tan(\frac{\pi J_z}{2})},
 \quad \lambda_\pi=\frac{\cot(\frac{\pi g}{2})}{\tan(\frac{\pi J_z}{2})}, 
\end{align}
of the MZMs and MPMs, respectively. Note that $\lambda_{0}$ and $\lambda_{\pi}$ are eigenvalues of the corresponding transfer matrix. This definition therefore differs from that used in Ref.\ \cite{Yeh2024}.

The correlation function $G_X(t)$ of the product of the Majorana modes clearly exhibits slow hybridization oscillations with two distinct periods, see lower panel of Fig.\ \ref{fig1}(d). Consistent with the product nature of the correlation function, these
periods $2\pi/\delta_{\pm}$ can be identified with $\delta_\pm=\delta_0\pm \delta_\pi$. We note that in Fig.\ \ref{fig1}(d), the transverse field $g$ deviates only slightly from the line $g=0.5$ along which $\xi_0=\xi_\pi$. This explains why the two periods of $G_X(t)$ are rather different in magnitude in Fig.\ \ref{fig1}(d). 

We can interpret the oscillation periods $\delta_{0,\pi}$ and $\delta_\pm$ in terms of the hybridization splittings of the spectral quadruplets, see Fig.\ \ref{figsplit}(a). Dropping the quadruplet index $n$ for simplicity, we have
\begin{align}
    E_{(1,n_\pi)}-E_{(0,n_\pi)}&=\delta_0,
    \notag
    \\
      E_{(n_0,1)}-E_{(n_0,0)}&=\pi+\delta_\pi,
    \label{eq:energy diff 0/pi}
\end{align}
for states differing in one of the Majorana occupations. Similarly, we have 
\begin{align}
    E_{(1,1)}-E_{(0,0)}&=\pi +\delta_+,
    \notag
    \\
    E_{(1,0)}-E_{(0,1)}&=\pi +\delta_-
     \label{eq:energy diff +-}
\end{align}
for states differing in both Majorana occupations. It is interesting to relate the splittings to fermion parity. In the fermionic formulation, the spin-flip symmetry maps to fermion parity $P=\prod_j e^{i\pi c_j^\dagger c_j}$. Each quadruplet falls apart into two pairs of states, a pair of one fermion parity consisting of $\ket{(0,0)}$ and  $\ket{(1,1)}$ and a pair of the other fermion parity consisting of $\ket{(1,0)}$ and  $\ket{(0,1)}$. We can thus identify $\delta_{0,\pi}$ as inter-parity splittings and $\delta_\pm$ as intra-parity splittings. This is reflected in that $G_Z(t)$ (with oscillation periods $\delta_{0,\pi}$) involves an operator that is odd under $P$, so that $E_n$ and $E_m$ in Eq.\ \eqref{eq:corrfunc} belong to states with different parities. In contrast, $G_X(t)$ (with oscillation periods $\delta_\pm$) involves an even operator, so that $E_n$ and $E_m$ in Eq.\ \eqref{eq:corrfunc} belong to states with identical parities. 

\section{Integrability-breaking coupling}
\label{sec:BreakIntegrability}

In the integrable case, inter- and intra-parity splittings are linearly dependent. Correspondingly, the slow oscillation periods of $G_Z(t)$ and $G_X(t)$ are of comparable magnitude. It was the main result of Ref.\ \cite{Yeh2024} that the two correlation functions exhibit very different sensitivities to the perturbation
\begin{align}
   V&=\frac{\pi J_x}{2} \sum^N_{j=1} X_jX_{j+1}
   \nonumber\\
   &= \frac{\pi J_x}{2} \sum^N_{j=1} (1-2c^\dagger_jc_j)(1-2c^\dagger_{j+1}c_{j+1}),
   \label{eq:pert x}
\end{align}
in the Floquet operator in Eq.\ \eqref{eq:UF}. This $J_x$-coupling maps to a four-fermion interaction and thus breaks integrability, but preserves the spin-flip (fermion-parity) symmetry $P$. 
The qualitatively different responses of $G_{X,Z}(t)$ to the $J_x$-coupling can be anticipated by considering the sweet spot $g=0.5$ and $J_z=1$. In fact, one notes that the MZM and MPM operators have a nonzero commutator with the perturbation,
\begin{align}
    [\gamma_{0,L},V]&=-i\pi J_x\gamma_{\pi,L}X_2,
    \label{eq:comm zero mode V}
    \\
    [\gamma_{\pi,L},V]&=i\pi J_x\gamma_{0,L}X_2.
    \label{eq:comm pi mode V}
\end{align}
Thus, the relations in Eq.\ \eqref{eq:comm 0 pi mode} no longer hold, which rapidly modifies $G_{Z}(t)$ with increasing $J_x$. In contrast, the product mode $\chi_L$ commutes with the perturbation
\begin{align}
    [\chi_{L},V]=0,
    \label{eq:comm prod mode V}
\end{align}
implying that Eq.\ \eqref{eq:comm prod mode} remains unchanged. As a result, the perturbation $V$ also leaves $G_X(t)$ unchanged at the sweet spot. In the following, we explore the dynamics away from the sweet spot.
\subsection{Numerical results}
\label{sec:numerical}
Figure \ref{fig3} shows late-time dynamics of the correlation functions for nonzero $J_x$. 
For $G^\pm_Z(t)$ [see Fig.\ \ref{fig3}(a)], we find that even small couplings $J_x$ suppress the hybridization-induced oscillations. The characteristic time beyond which the correlation functions $G^\pm_Z(t)$ decay, becomes rapidly smaller than the inverse Majorana splittings of the unperturbed spectrum. Moreover, the behaviors of the two correlation functions $G^\pm_Z(t)$ become essentially identical. 

This is reflected in the corresponding Fourier transforms $G^\pm_Z(\omega)$ [Fig.\ \ref{fig3}(b)]. In the integrable limit ($J_x=0$), the Fourier transform exhibits peaks at the inter-parity splittings $\delta_0$ for $G^+_Z(\omega)$ and $\delta_\pi$ for $G^-_Z(\omega)$. We note that the broadening of the peaks for $J_x=0$ arises from the finite simulation time. A small perturbation with $J_x \sim 10^{-5}$ essentially suppresses these peaks, consistent with the rapid suppression of the hybridization-induced oscillations in $G_Z(t)$ with $J_x$.  

In contrast, long-time traces of $(-1)^tG_X(t)$ exhibit  hybridization-induced splitting oscillations up to much larger couplings $J_x$ [Fig.\ \ref{fig3}(d)]. We also observe that the superposition of  oscillations with different periods (beating) survives significantly larger perturbations. 
The corresponding Fourier transform  $G_X(\omega)$ [Fig.\ \ref{fig3}(e)] features two peaks close to $\omega\simeq \pi$, which progressively broaden with increasing coupling strength $J_x$. The peaks derive from the oscillation periods $\delta_\pm$ in the limit $J_x=0$ and remain distinguishable up to couplings $J_x\lesssim 10^{-3}$. This is two orders of magnitude larger than for $G^\pm_Z(\omega)$.  

\begin{figure}[t!]
    \centering
    \includegraphics[width=\linewidth]{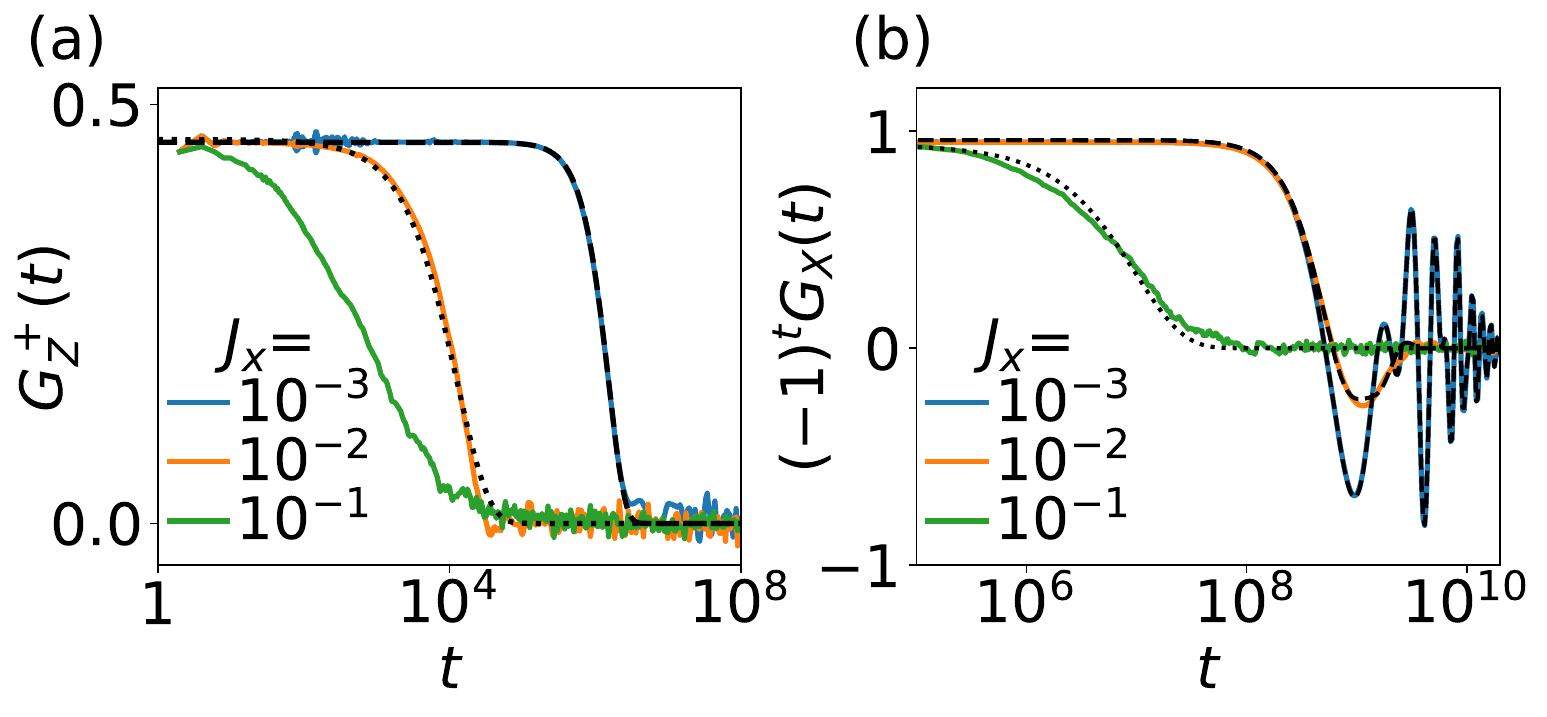}
    \caption{Temporal decay of correlation functions in the presence of coupling $J_x$ (data: colored lines). (a) $G^+_Z(t)$ decays as a Gaussian (black dashed) for intermediate couplings and as an exponential for larger couplings $J_x$ (black dotted). (b)  $G_X(t)$ decays as a linear combination of terms with Gaussian envelope as detailed in Eq.\ \eqref{eq:fit double Gauss} (black dashed) for intermediate couplings and as an exponential for larger interactions (black dotted). Parameters: $N=12$, $J_z=0.9$, $g=0.52$.}
    \label{fig4}
\end{figure}

We fit the decay behaviors of $G^+_Z(t)$ and $G_X(t)$ in Fig.\ \ref{fig4}. For $G^+_Z(t)$, we find good agreement with a Gaussian fit for coupling strengths $J_x\lesssim 10^{-3}$ [black dashed line in Fig.\ \ref{fig4}(a)], and an exponential fit for larger $J_x$ (dotted line). For $G_X(t)$, the two oscillation periods motivate a fit using a superposition of two terms with Gaussian envelopes,
\begin{align}
    G_X(t)\simeq A(-1)^t\left\{ e^{-t^2/\tau^2_+}\cos(\delta_+t)+e^{-t^2/\tau^2_-}\cos(\delta_-t) \right\}
    \label{eq:fit double Gauss}. 
\end{align}
Here, $\tau_\pm$ are  characteristic time scales  and  $A$ is a prefactor. We find excellent agreement with the Gaussian fit function in Eq.\ \eqref{eq:fit double Gauss} [dashed line in Fig.\ \ref{fig4}(b)], up to larger coupling strengths $J_x\lesssim J_\times \sim 10^{-2}$ than for $G_Z(t)$. The correlation function decays exponentially for larger couplings $J_x$ (dotted line).

\begin{figure*}[t!]
    \centering
    \includegraphics[width=\linewidth]{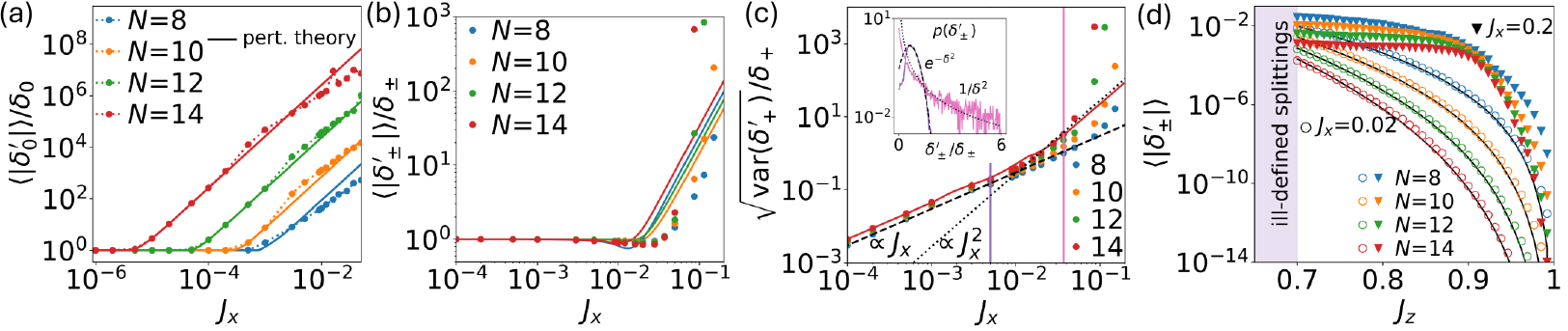}
    \caption{Splittings in the presence of transverse couplings $J_x$. (a) Typical inter-parity splittings are lifted by small $J_x$ (double-logarithmic scale). Dash-dotted: exact-diagonalization data. Solid: perturbation theory. (b)  Typical intra-parity splittings are robust against large $J_x$ (semi-logarithmic scale). Dots: exact-diagonalization data. Solid: perturbation theory. (c) 
    Intra-parity splitting fluctuations for which $|\delta^\prime_{+}|>\delta_{t}$. The threshold $\delta_{t}=(\delta_++\delta_-)/2$ is chosen to separate the peaks in the intra-parity splitting distributions. Solid: Perturbation theory. Black dashed and black dotted lines indicate linear ($\propto J_x$) and quadratic ($\propto J_x^2$) scaling, respectively, and serve as guides to the eye for identifying the relevant perturbative order. Inset: Distributions at the corresponding $J_x$-couplings (see colors), with respective Gaussian and Lorentzian fits.
    (d) Typical intra-splittings vs.\ $J_z$-coupling. Solid black lines for bare splittings. Parameters: (a)-(d) $g=0.52$; (a)-(c) $J_z=0.9$.}
    \label{fig5}
\end{figure*}

We relate the different behaviors of the correlation functions to the level statistics of the spectral quadruplets. We compute the inter- and intra-parity splittings 
\begin{align}
  E^{e}_n-E^o_{n}&=\delta^\prime_{n,0},
  \\
  E^{e}_n-E^o_{n+2^{N-2}}&=\pi+\delta^\prime_{n,\pi},
  \\
    E^{e(o)}_n-E^{e(o)}_{n+2^{N-2}} &= \pi+\delta^\prime_{n,\pm},
\end{align}
by exact diagonalization from the many-body spectrum, see Fig.\ \ref{figsplit}(b). Here, $E^{e(o)}_{n}$ are sorted eigenphases of the many-body Floquet operator for the even ($e$) and odd ($o$) fermion parity sectors. The primes distinguish splittings in the presence of finite $J_x$ from those in the integrable limit, cf.\  Eqs.\ \eqref{eq:energy diff 0/pi} and \eqref{eq:energy diff +-}. Quadruplets can be identified as long as these splittings are small compared to the mean level spacing $(2\pi)/2^{N}$. For each quadruplet, there are up to three independent eigenphase differences. We take into account the four possible inter-parity splittings (two zero, two $\pi$), and two intra-parity $\pi$-splittings per quadruplet which introduces some redundancy. However, this parallels the appearance of all possible energy differences in the exact-eigenstate representation of the correlation function in Eq.\ \eqref{eq:corrfunc}. While the splittings are identical for all quadruplets in the integrable limit, they become nonuniform (i.e., dependent on the quadruplet $n$) under the integrability-breaking perturbation.  

The resulting splitting distributions, obtained by histogramming the splittings across the many-body spectrum, reveal very different behaviors in the presence of interactions $J_x$, see Fig.\ \ref{fig3}(c) and (f). The inter-parity splittings are affected for small $J_x \simeq 10^{-5}$ [Fig.\ \ref{fig3}(c)]. The distributions $p(\delta_0^\prime)$ and  $p(\delta_\pi^\prime)$ flatten out and become progressively identical. In contrast, the intra-parity splittings are only affected for much larger interactions $J_x\gtrsim 10^{-3}$ [Fig.\ \ref{fig3}(f)]. The double-peak structure of the distributions $p(\delta^\prime_\pm)$ derived from the unperturbed splittings $\delta_\pm$ broadens. Nevertheless, a merged peak in the splitting distribution remains present for values $J_x\gtrsim 10^{-2}$.  

Remarkably, the splitting distributions in Fig.\ \ref{fig3}(c) and (f) are closely related to the Fourier transforms of the correlation functions in Fig.\ \ref{fig3}(b) and (e). In the comparison, one should remember that the splitting distribution is completely unbroadened in the integrable limit, unlike the peaks in $G_Z(\omega)$ and $G_X(\omega)$ which broaden due to the finite simulation time. The correspondences between Fourier transforms of correlation functions and splitting distributions can be understood from the exact-eigenstate expression for the correlation function in Eq.\ \eqref{eq:corrfunc}. Since $Z_1$ is odd under the Ising symmetry, the exact eigenstates $E_n$ and $E_m$ entering the correlation function $G_Z(t)$ have opposite fermion parities. Thus, its hybridization-induced oscillations emerge from terms in the sum Eq.\ \eqref{eq:corrfunc}, in which $E_n$ and $E_m$ are part of the same quadruplet and split by $\delta_0^\prime$ or $\delta_\pi^\prime$. Correspondingly, the Fourier transform $G_Z(\omega)$ is proportional to the distribution of inter-parity splittings, provided that the fluctuations of the matrix elements can be neglected. The same line of arguments holds for $G_X(t)$, except that $X_1$ is even under the spin-flip symmetry, so that $E_n$ and $E_m$ have the same fermion parity. Thus, the Fourier transform $G_X(\omega)$ is proportional to the distribution of intra-parity splittings.

Figure \ref{fig5} shows a finite-size analysis of mean splittings in the presence of interactions, averaged over the many-body spectrum. On the one hand, the mean inter-parity splitting (illustrated by $\langle |\delta^\prime_{0}|\rangle$) rapidly increases with growing $J_x$ above a critical coupling strength, see Fig.\ \ref{fig5}(a). Comparing various system sizes $N$, we deduce that the critical coupling depends exponentially on $N$, i.e., $J_{x}^{(\textrm{crit})} \simeq 2^{-N}$. This is consistent with the sensitivity of $G_Z(t)$ to small perturbations. On the other hand, the mean intra-parity splitting $\langle |\delta^\prime_{\pm}|\rangle$ remains approximately unaffected up to large couplings $J_x$, exhibiting only a weak dependence on $N$, see Fig.\ \ref{fig5}(b). This is consistent with the robustness of $G_X(t)$ to the integrability-breaking perturbation. Figure \ref{fig5}(d) maps the robustness of intra-parity splittings as a function of $J_z$, for finite and fixed $J_x$. We find that the mean intra-parity splittings remain close to the unperturbed value throughout the entire range $0.71\leq J_z\leq 1$, provided that $J_x\lesssim J_\times \sim 10^{-2}$. Notice that at $J_z\sim 0.71$, the unperturbed splittings approach the level spacing and cannot be spectrally resolved (``ill-defined splittings"). Near $J_z \sim 1$, the splittings remain well below the level spacing only for $J_x \lesssim 0.2$.

\begin{figure}[b!]
    \centering
\includegraphics[width=0.98\linewidth]{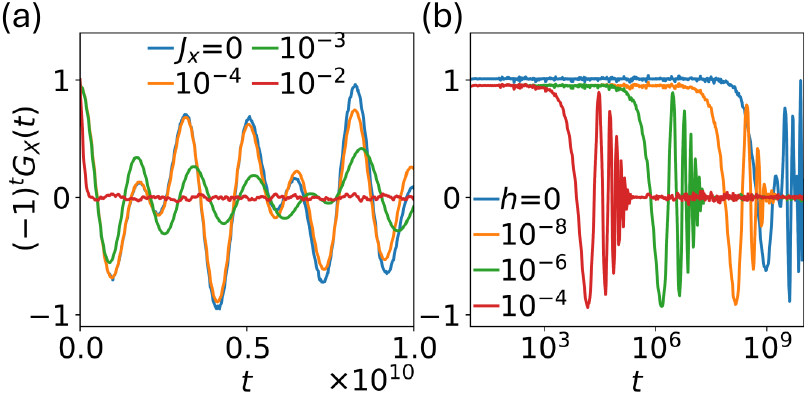}
    \caption{Integrability vs.\ symmetry breaking. (a) $G_X(t)$ remains long-lived in the presence of  additional next-nearest-neighbor couplings, which break integrability but preserve the spin-flip symmetry, see Eq.\ \eqref{eq:integrability breaking} and parameters below. (b) $G_X$ relaxes quickly for interactions which break both integrability and symmetry, see Eq.\ \eqref{eq:symmetry breaking}. Parameters: $N=12$, $g=0.52$, $J_z=0.9$, (a) $J_x=J_{x,2}=J_{z,2}$, (b) $J_x=0$.}
    \label{fig:integ vs symmetry breaking}
\end{figure}

It is interesting to investigate the stability of the product mode for more general couplings. We first modify the Floquet operator by including next-nearest-neighbor exchange, 
\begin{align}
    U_F \rightarrow 
U_F\,  e^{\frac{i\pi J_{x,2}}{2}\sum^{N-2}_{j=1}X_jX_{j+2}}e^{\frac{i\pi J_{z,2}}{2}\sum^{N-2}_{j=1}Z_jZ_{j+2}}.
\label{eq:integrability breaking}
\end{align}
The additional couplings break integrability but preserve the spin-flip symmetry. Figure \ref{fig:integ vs symmetry breaking}(a) shows dynamics of the correlation function $G_X(t)$, which is qualitatively similar to the dynamics in the absence of the additional terms as shown in Fig.\ \ref{fig3}(d). 

In contrast, interactions which break both integrability and spin-flip symmetry induce significant relaxation of the product mode. Figure \ref{fig:integ vs symmetry breaking}(b) shows results for a modified Floquet operator, which includes a longitudinal field,  
\begin{align}
    U_F \rightarrow 
U_F\,  e^{\frac{i\pi h}{2}\sum^{N}_{j=1}Z_j}.
\label{eq:symmetry breaking}
\end{align}
The longitudinal field couples quadruplet states of opposite parity. As such, the product mode is less stable than $\pi$ modes in the MPM phase, which are very robust against longitudinal fields \cite{Schmid2024}.    

\subsection{Analytical considerations}
\label{sec:analytical}

The contrasting behaviors of intra- and inter-parity splittings -- and thus correlation functions -- can be traced to the fact that the $J_x$-coupling breaks integrability but conserves the global spin-flip symmetry. Levels within a single symmetry sector repel, while levels from different symmetry sectors cross freely. This distinction implies that the levels of an inter-parity pair evolve independently governed only by level repulsion in their respective parity sectors. Conversely, intra-parity $\pi$-pairs evolve in a correlated manner. 

We consider inter- and intra-parity splittings in Floquet perturbation theory. In the absence of degeneracies, we can expand the eigenphases of $U_F=U_{F,0}e^{-iV}$ as $E_n = E_{n,0}+ E_{n,1} +  E_{n,2} + \ldots$ with \cite{Schmid2024,Penner2025}
\begin{equation}
     E_{n,1} = 
     \matrixel{n_0}{V}{n_0}
   \,\,\, ; \,\,\,  
     E_{n,2} =\sum_{m \neq n} \frac{|\matrixel{n_0}{V}{m_0}|^2}{2\tan \frac{E_{n,0}-E_{m,0}}{2}}. 
     \label{eq:pt}
\end{equation}
The eigenphase denominator is small for nearby levels and diverges for  antipodal levels with  $E_{n,0}-E_{m,0} \simeq \pm \pi$. We can use the perturbation theory to expand the inter- and intra-parity  splittings to second order in $V$,
 \begin{align}
\delta^\prime_{n,0}&\simeq \delta_{0} + \delta^{(1)}_{n,0} + \delta^{(2)}_{n,0} 
\\
 \delta^\prime_{n,+} & \simeq \delta_+ + \delta^{(1)}_{n,+} + \delta^{(2)}_{n}.
\end{align}
The order in perturbation theory is indicated by the superscript. Analogous expansions hold for  $\delta'_{n,\pi}$ and $\delta'_{n,-}$.

Inter-parity splittings are dominated by the second-order contribution, with the first order being negligible. This can be inferred from the slopes in Fig.\ \ref{fig5}(a). Accordingly, we find that the inverse decay time of $G_Z(t)$ grows quadratically with coupling strength. In contrast, intra-parity splittings remain largely unaffected up to $J_x\lesssim J_\times$ [Fig.\ \ref{fig5}(b)], and increase drastically beyond this coupling. This behavior indicates that they remain essentially  unaffected in low-order perturbation theory. Our goal here is to gain analytical insight into these observations    based on perturbation theory.

We begin by examining the first-order terms which are expected to be small in both cases. It is useful to write them as commutators 
\begin{align}
\delta^{(1)}_{n,0}&=\mel{n,00}{V}{n,00}-\mel{n,10}{V}{n,10}
\notag
\\
&=\bra{n,00}\big[\gamma_{0,L},V \big]\gamma_{0,L}\ket {n,00},
\label{eq:1st order pt delta0}
\\
\delta^{(1)}_{n,+}&=\mel{n,00}{V}{n,00}-\mel{n,11}{V}{n,11}
\notag
\\
&=\bra{n,{00}}\big[\chi_{L},V \big]\chi_{L}\ket {n,{00}}.
\label{eq:1st order pt deltaplus}
\end{align}
The matrix elements can be estimated by observing that the perturbation $V$ commutes with all $X_j$ operators, and that near the sweet spot one has $X_1 \simeq i \gamma_{0,L}\gamma_{\pi,L}=\chi_L$ up to exponentially small corrections in $N$. Consequently, the commutator in $\delta^{(1)}_{n,+}$ is exponentially small. In contrast, the commutator entering $\delta^{(1)}_{n,0}$ is in general nonzero, in accordance with Eq.\ \eqref{eq:comm zero mode V}. However, its expectation value $\delta^{(1)}_{n,0}\propto \bra{n,00}X_2\gamma_{0,L}\gamma_{\pi,L}\ket {n,00}$, which enters first-order perturbation theory, is exponentially small. We also verify this numerically (see also App.\ \ref{app:matrix elements}).

We now turn to the second-order corrections to the intra-parity splittings. Within a given quadruplet, the perturbation acts only between equal-parity levels, e.g., $\ket{n,00}$ and $\ket{n,11}$. As these are $\pi$-paired, the corresponding contribution to second-order perturbation theory can be neglected. We can thus restrict attention to inter-quadruplet contributions,
\begin{align}
    \delta^{(2)}_{n,+}=
    &\sum_{ m\neq n}
    \left\{\frac{|\langle n, 11| V | m,11\rangle|^2}{2\tan\frac{E_{n,11}-E_{m,11}}{2}}
    -\frac{|\langle n, 00| V | m,00\rangle|^2}{2\tan\frac{E_{n,00}-E_{m,00}}{2}}\right\}
    \notag
    \\
    +&\sum_{ m\neq n}
    \left\{\frac{|\langle n,11| V | m,00\rangle|^2}{2\tan\frac{E_{n,11}-E_{m,00}}{2}}
    -\frac{|\langle n, 00| V | m,11\rangle|^2}{2\tan\frac{E_{n,00}-E_{m,11}}{2}}\right\}
    \notag
    \\
    +&\sum_{ m\neq n}
    \left\{\frac{|\langle n, 11| V | m,01\rangle|^2}{2\tan\frac{E_{n,11}-E_{m,01}}{2}}
    -\frac{|\langle n, 00| V | m,10\rangle|^2}{2\tan\frac{E_{n,00}-E_{m,10}}{2}}\right\}
     \notag
    \\
    +&\sum_{ m\neq n}
    \left\{\frac{|\langle n, 11| V | m,10\rangle|^2}{2\tan\frac{E_{n,11}-E_{m,10}}{2}}
    -\frac{|\langle n, 00| V | m,01\rangle|^2}{2\tan\frac{E_{n,00}-E_{m,01}}{2}}\right\}. 
    \label{eq:pert theory 2nd order} 
\end{align}
In the first two lines, the matrix elements involve states which differ in the occupations of none or both of the end modes (and hence an even number of bulk quasi-particles). In the last two lines, they differ by a MZM or a MPM (and an odd number of bulk quasi-particles). The energy denominators of the terms subtracted in each line are either exactly equal, e.g., $E_{n,11}-E_{m,11}=E_{n,00}-E_{m,00}$, or equal up to an exponentially small (unperturbed) splitting, e.g., $E_{n,11}-E_{m,01}=E_{n,00}-E_{m,10}+\delta_0$.

\setlength{\tabcolsep}{0pt}
\begin{table}[t!]
    \centering
    \begin{tabular}{|c||c|c|c|c|}\hline
        & MZM & MPM & both & none \\\hline\hline
        \multicolumn{5}{|c|}{$\delta_\pm$} \\\hline
        energy denom. & $\pm \pi/2$ & $\pm \pi/2$ & $0,\pi$ & $0,\pi$ \\\hline
        matrix elem.\ diff. & $\neq 0$ & $\neq 0$ & 0 & 0 \\\hline        
        $\;\;$matrix elem.\ size$\;\;$ & $\;\;$small$\;\;$ & $\;\;$small$\;\;$ & $\;\;$large$\;\;$ & $\;\;$large$\;\;$ \\\hline\hline
        \multicolumn{5}{|c|}{$\delta_{0}$} \\\hline
        energy denom. & $\pm \pi/2$ & $\pm \pi/2$ &\cellcolor{yellow}$0,\pi$ & $0,\pi$ \\\hline
        matrix elem.\ diff. & $\neq 0$ & 0 &\cellcolor{yellow}$\neq 0$ & 0 \\\hline
        matrix elem.\ size & small & small &\cellcolor{yellow}large & large \\\hline
    \end{tabular}
    \caption{Modulus squared of the matrix elements as well as energy denominators for the different contributions to second-order perturbation theory. Columns refer to the different lines in Eqs.\ \eqref{eq:pert theory 2nd order} and \eqref{eq:2nd order inter-parity}.
    A zero matrix element difference indicates that it is exponentially small in $N$. The energy denominators can be zero up to  polynomially small corrections in $N$. While none of the contributions are sizable for the intra-parity splitting, the inter-parity splitting $\delta_{0,\pi}$ is sizable due to the contribution in the ``both" column (yellow mark).}
    \label{tab1}
\end{table}

Near the sweet spot, all terms in Eq.\ \eqref{eq:pert theory 2nd order} are small due to either large energy denominators or small matrix elements, see Tab.\ \ref{tab1}. Terms in the last two lines have energy denominators  $E_{n,11}-E_{m,10}\simeq \pm \pi/2$. To see this we note that near the sweet spot, the bulk quasiparticles have single-particle energies $\approx \pi/2$, see Eq.\ \eqref{eq:excitation spectrum}. The magnitude of the contributions is therefore controlled by the differences of matrix elements. We find them to be numerically small (App.\ \ref{app:matrix elements}). Terms in the first two lines may have \textit{polynomially} small energy denominators  $E_{n,11}-E_{m,10}\simeq 0$. However, the matrix element differences are \textit{exponentially} small. The latter can be inferred from symmetrizing the matrix elements and expressing them via (anti-)commutators, e.g.,
\begin{align}
&\bra{n,11}V\ket{m,00}\pm  \bra{n,00}V\ket{m,11} 
\notag
\\
=&-i\bra{n,00}[\chi_{L},V]_{\mp}\ket{m,00}.
\end{align}
We also confirm numerically that intra-parity splittings are exponentially small in second order (App.\ \ref{app:matrix elements}). 

We can use the perturbative approach to gain insight into the time dependence of $G_X(t)$ shown in Fig.\ \ref{fig4}(b). We plot the variance of the intra-parity splittings along with representative full distributions (inset) in Fig.\ \ref{fig5}(c). We observe that the variance is governed by the first-order terms for $J_x\lesssim J_\times \sim 10^{-2}$. For  $J_x$ larger than the crossover scale $J_\times$, both the average [Fig.\ \ref{fig5}(b)] and the variance [Fig.\ \ref{fig5}(c)] rapidly grow due to higher-order terms in the perturbation expansion. The first-order contribution to the splitting is a sum over many random terms. This results in a Gaussian splitting distribution and hence a Gaussian decay of the boundary spin correlations $G_X(t)$. For larger $J_x$, the splittings are dominated by  higher-order terms in the perturbation expansion. These are also sums over many terms. Importantly, they involve energy denominators in addition to matrix elements. The sum will typically be dominated by terms with small energy denominators. As a result, these sums are Levy flights and do not obey the conventional central limit theorem \cite{Bouchaud1990}. One readily argues that assuming statistically independent terms, the stable distribution of the sum is a Lorentzian, see e.g., \cite{Penner2021}. This directly translates into  an exponential decay of the boundary spin correlations $G_X(t)$ and therefore explains the crossover between the Gaussian and exponential dependencies of $G_X(t)$, which we observe numerically. A similar picture arises in the MPM phase in the presence of symmetry-breaking perturbations \cite{Schmid2024}.
We note that numerically, the crossover scale $J_\times$ seems to depend only weakly on the length of the chain. 

Finally, we consider the second-order correction to the inter-parity splittings,
\begin{align}
  \delta^{(2)}_{n,0}=&\sum_{ m\neq n}
    \left\{\frac{|\langle n,10| V | m,10\rangle|^2}{2\tan\frac{E_{n,10}-E_{m,10}}{2}}
    -\frac{|\langle n,00| V | m,00\rangle|^2}{2\tan\frac{E_{n,00}-E_{m,00}}{2}}\right\}  
    \notag
    \\
    +& \sum_{ m\neq n}
    \left\{\frac{|\langle n,10| V | m,01\rangle|^2}{2\tan\frac{E_{n,10}-E_{m,01}}{2}}
    -\frac{|\langle n,00| V | m,11\rangle|^2}{2\tan\frac{E_{n,00}-E_{m,11}}{2}}\right\} 
    \notag
    \\
    +&\sum_{ m\neq n}
    \left\{\frac{|\langle n,10| V | m,00\rangle|^2}{2\tan\frac{E_{n,10}-E_{m,00}}{2}}
    -\frac{|\langle n,00| V | m,10\rangle|^2}{2\tan\frac{E_{n,00}-E_{m,10}}{2}}\right\}
    \notag
    \\
    +&\sum_{ m\neq n}
    \left\{\frac{|\langle n,10| V | m,11\rangle|^2}{2\tan\frac{E_{n,10}-E_{m,11}}{2}}
    -\frac{|\langle n,00| V | m,01\rangle|^2}{2\tan\frac{E_{n,00}-E_{m,01}}{2}}\right\}.
    \label{eq:2nd order inter-parity}
\end{align}
Contributions are grouped by the number of end modes involved: none, both, MZM, or MPM. Unlike the intra-parity splittings, the contribution to the inter-parity splitting of the second line 
in Eq.\ \eqref{eq:2nd order inter-parity} can be large, see highlighted column labeled ``both" in Tab.\ \ref{tab1}. 
The corresponding term can have small energy denominators $E_{n,10}-E_{m,01}\simeq 0$, while the associated matrix elements are appreciable, i.e., not exponentially small. The matrix elements can be inferred by symmetrizing and expressing them in terms of (anti-)commutators,
\begin{align}
&\bra{n,10}V\ket{m,01}\pm  \bra{n,00}V\ket{m,11} 
\notag
\\
& \qquad =\bra{n,00}[\gamma_{0,L},V]_{\pm}\gamma_{\pi,L}\ket{m,00},
\end{align}
which is not suppressed. In contrast, all other contributions are subdominant, either because of larger energy denominators $\sim \pi/2$ or because of small matrix elements, see Tab.\ \ref{tab1} and App.\ \ref{app:matrix elements}. 

The statistics of inter-parity splittings crosses over from Gaussian to Lorentzian when small energy denominator  become relevant in the second-order terms. Here, this crossover occurs for exponentially small couplings in $N$, which explains the sensitivity of $G_Z(t)$ [Fig.\ \ref{fig4}(a)].

\section{Conclusions}
\label{sec:conclusions}

We investigated strong-mode physics in the $0\pi$ phase of the Floquet quantum Ising model. We showed that the different levels of robustness of the individual Majorana modes compared to the Floquet product mode can be illuminated from the point of view of spectral pairings. The coexistence of two types of Majorana operators implies that in the integrable limit, the eigenphases appear in quadruplets. Under integrability-breaking perturbations, these quadruplets necessarily fall apart into two doublets. These doublets are characterized by the eigenvalues under the protecting $\mathbb{Z}_2$ symmetry. While there are strong spectral correlations within the symmetry sectors, states in different symmetry sectors are essentially uncorrelated. This is the reason underlying the hierarchy of strong modes in the $0\pi$ phase. The individual MZMs and MPMs couple levels from different symmetry sectors, leading to limited stability against integrability-breaking perturbations. In contrast, the product mode of both Majorana operators couples states from the same symmetry sector. The spectral rigidity within the symmetry sectors enhances the stability against 
integrability-breaking perturbations. 

It is an interesting question for future research whether similar hierarchies of strong modes appear in other models. Our work suggests that there can be a finetuned (e.g., integrable) limit, in which the eigenphase order involves groups of levels, which is larger than the number of distinct eigenvalues of the symmetry operator. Product edge modes should generically exist when multiple edge modes localize at the same end of a system. One interesting class of models are periodically driven clock models, which host up to $n$ parafermionic edge modes dictated by their $\mathbb{Z}_n$-symmetry \cite{Fendley2012,Jermyn2014, Fendley2014,Moessner2016,Alicea2016,Surace2019}. Thus, these models  present an interesting platform for constructing  multiple composite edge objects that may remain stable in the presence of integrability-breaking perturbations. Future work could also consider product modes beyond periodic driving schemes such as quasi-periodic \cite{Verdeny2016,Dumitrescu2018,Peng2018,Crowley2019,Maity2019,Lapierre2020,Long2021,Bhattacharjee2022,Friedman2022,Schmid2024a}, random multipolar \cite{Zhao2021,Zhao2022,Zhao2023,Liu2025} or non-unitary driving \cite{Wen2021,Lapierre2025a,Lapierre2025}.  

\begin{acknowledgments}
We thank A.\ Mitra and A.-G.\ Penner for useful discussions. We gratefully acknowledge funding by Deutsche Forschungsgemeinschaft through CRC 183. The bachelor  studies of one of us (FM) were supported by a scholarship of the Studienstiftung des dt.\ Volkes. We thank the HPC
service of ZEDAT, Freie Universität Berlin, for computing
time \cite{Bennett2020}.
\end{acknowledgments}

\appendix

\section{Numerical analysis of the perturbative splittings}
\label{app:matrix elements}

\begin{figure}[h!]
    \centering
\includegraphics[width=0.99\linewidth]{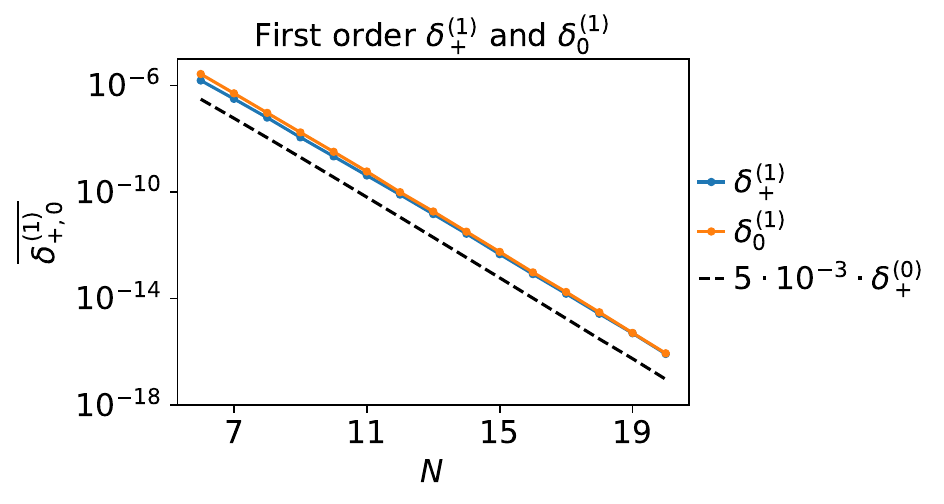}
    \caption{First-order perturbative correction to intra- and inter-parity splittings. Parameters: $J_z=0.9$, $g=0.52$, $J_x=0.001$.}
    \label{fig:1st_order_comparison}
\end{figure}

\begin{figure}[h!]
    \centering
\includegraphics[width=0.99\linewidth]{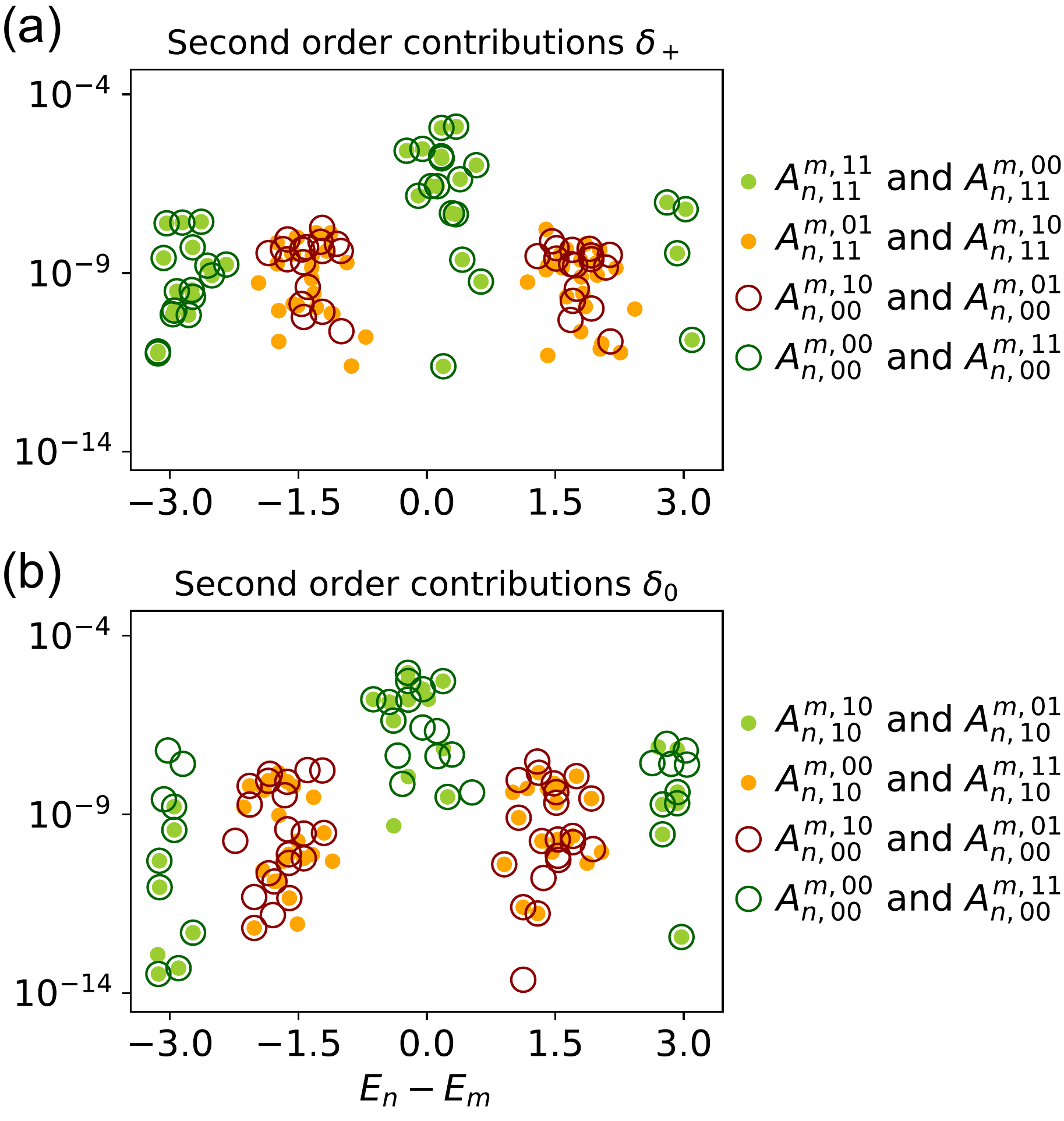}
    \caption{
    Individual terms of the  second order perturbative corrections to the intra- and inter-parity splittings for the different processes, see Eqs.\ \eqref{eq:corr_deltaplus}, \eqref{eq:corr_delta0}. (a) Intra-parity splittings. (b) Inter-parity splitings. Same parameters as in Fig.\ \ref{fig:1st_order_comparison} and $N=9$.
    }
    \label{fig:me_contributions_new}
\end{figure}

\begin{figure*}[t!]
    \centering
\includegraphics[width=\linewidth]{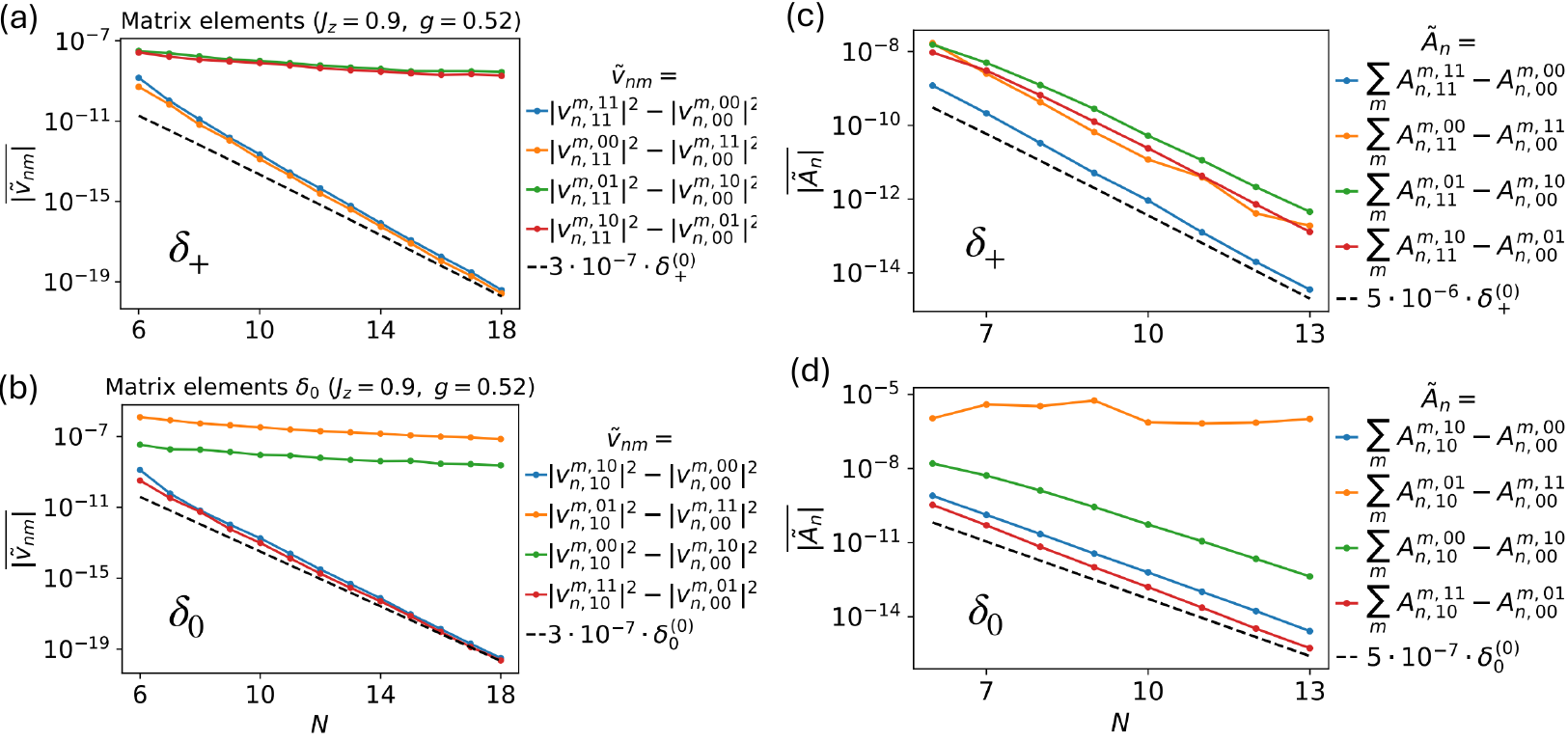}
    \caption{Second order perturbation theory. (a),(b)  spectrally averaged differences of the
modulus squared of matrix elements for intra-parity splittings and inter-parity splittings. (c),(d) Corresponding total contributions. Parameters: $J_z=0.9$, $g=0.52$, $J_x=0.001$}
    \label{fig:app}
\end{figure*}

We numerically compute the various contributions to the intra- and inter-parity splittings appearing in perturbation theory. Figure \ref{fig:1st_order_comparison} shows the spectral average of the first-order corrections. We find that both intra- and inter-parity splittings decay exponentially with system size. 

For the second-order corrections, it is useful to write Eq.\ \eqref{eq:pert theory 2nd order} as
\begin{align}
\delta_{n,+}^{(2)} &= \sum_{m\neq n}\big(A_{n,11}^{m,11}-A_{n,00}^{m,00}\big)
    +\sum_{m\neq n}\big(A_{n,11}^{m,00}-A_{n,00}^{m,11}\big)
    \notag
    \\
    &+  \sum_{m\neq n}\big(A_{n,11}^{m,01}-A_{n,00}^{m,10}\big)
    +\sum_{m\neq n}\big(A_{n,11}^{m,10}-A_{n,00}^{01}\big)
    \label{eq:corr_deltaplus}
\end{align}
and Eq.\ \eqref{eq:2nd order inter-parity} as
\begin{align}
\delta_{n,0}^{(2)} &= \sum_{m\neq n}\big(A_{10}^{m,10}-A_{n,00}^{m,00}\big)
    +\sum_{m\neq n}\big(A_{n,10}^{m,01}-A_{n,00}^{m,11}\big)
    \notag
    \\
    &+  \sum_{m\neq n}\big(A_{n,10}^{m,00}-A_{n,00}^{m,10}\big)
    +\sum_{m\neq n}\big(A_{n,10}^{m,11}-A_{n,00}^{01}\big).
    \label{eq:corr_delta0}
\end{align}
We have defined the individual contributions
\begin{align}
    A_{n,n_0 n_\pi}^{m,m_0 m_\pi} &= \frac{|v^{m,m_0 m_\pi}_{n,n_0 n_\pi}|^2}{2\,\tan\frac{E_{n,n_0 n_\pi}-E_{m,m_0 m_\pi}}{2}}
\end{align}
and abbreviated the matrix elements
$v^{m,m_0 m_\pi}_{n,n_0 n_\pi} = \mel{m,m_0m_\pi}{V}{n,n_0n_\pi}$.

We first analyze the intra-parity splittings. 
Figure \ref{fig:me_contributions_new}(a) shows a scatter plot of the individual terms $A$ in Eq.\ \eqref{eq:corr_deltaplus} for a representative pair of states $n$. The two terms within one sum in the equation are plotted as full and empty circles, respectively. Contributions from states which differ in none or both of the end modes (green dots and circles) come with energy denominators $\approx 0,\pi$. While the ones with zero energy denominators are substantially larger, the two terms contributing to the corresponding sums in Eq.\ \eqref{eq:corr_deltaplus} nearly cancel for individual values of $m$, as green dots and circles systematically coincide. In contrast, contributions from states which differ by a MZM or MPM (orange dots and red circles) come with energy denominators $\approx \pm \pi/2$. The corresponding subtracted terms do not cancel individually, but are systematically small.

Figure \ref{fig:app}(a) shows spectrally averaged differences of the modulus squared of matrix elements. We find that differences for states which differ by none or both of the end modes decay exponentially with system size. Their behavior essentially follows the decay of $\delta_+$ (black dashed line). Conversely, differences for states which differ by a MZM or MPM depend only weakly on $N$. Figure \ref{fig:app}(c) shows the corresponding full contributions to second-order perturbation theory, including energy denominators and sums over $m$. We find that these are exponentially suppressed in $N$ for all four processes. 

Next, we analyze the inter-parity splittings in second order. Figure \ref{fig:me_contributions_new}(b)
shows that the individual terms $A$ in Eq.\ \eqref{eq:corr_delta0} no longer coincide in all cases, even for the large contributions with energy denominators close to zero.
The  spectrally averaged differences of the modulus squared of the matrix elements shown in Fig.\ \ref{fig:app}(b) decay exponentially in system size for states which differ by none or a MPM mode. In contrast, matrix element differences of states which differ by a MZM or both of the end modes depend only weakly on $N$. Finally, the total contributions to the second-order perturbation theory as shown in Fig.\ \ref{fig:app}(d)
remains sizable with increasing $N$ for the contribution which differs in both of the end modes. The other three processes are exponentially suppressed. 

%
%

\begin{thebibliography}{56}%
\makeatletter
\providecommand \@ifxundefined [1]{%
 \@ifx{#1\undefined}
}%
\providecommand \@ifnum [1]{%
 \ifnum #1\expandafter \@firstoftwo
 \else \expandafter \@secondoftwo
 \fi
}%
\providecommand \@ifx [1]{%
 \ifx #1\expandafter \@firstoftwo
 \else \expandafter \@secondoftwo
 \fi
}%
\providecommand \natexlab [1]{#1}%
\providecommand \enquote  [1]{``#1''}%
\providecommand \bibnamefont  [1]{#1}%
\providecommand \bibfnamefont [1]{#1}%
\providecommand \citenamefont [1]{#1}%
\providecommand \href@noop [0]{\@secondoftwo}%
\providecommand \href [0]{\begingroup \@sanitize@url \@href}%
\providecommand \@href[1]{\@@startlink{#1}\@@href}%
\providecommand \@@href[1]{\endgroup#1\@@endlink}%
\providecommand \@sanitize@url [0]{\catcode `\\12\catcode `\$12\catcode
  `\&12\catcode `\#12\catcode `\^12\catcode `\_12\catcode `\%12\relax}%
\providecommand \@@startlink[1]{}%
\providecommand \@@endlink[0]{}%
\providecommand \url  [0]{\begingroup\@sanitize@url \@url }%
\providecommand \@url [1]{\endgroup\@href {#1}{\urlprefix }}%
\providecommand \urlprefix  [0]{URL }%
\providecommand \Eprint [0]{\href }%
\providecommand \doibase [0]{https://doi.org/}%
\providecommand \selectlanguage [0]{\@gobble}%
\providecommand \bibinfo  [0]{\@secondoftwo}%
\providecommand \bibfield  [0]{\@secondoftwo}%
\providecommand \translation [1]{[#1]}%
\providecommand \BibitemOpen [0]{}%
\providecommand \bibitemStop [0]{}%
\providecommand \bibitemNoStop [0]{.\EOS\space}%
\providecommand \EOS [0]{\spacefactor3000\relax}%
\providecommand \BibitemShut  [1]{\csname bibitem#1\endcsname}%
\let\auto@bib@innerbib\@empty
\bibitem [{\citenamefont {Fendley}(2012)}]{Fendley2012}%
  \BibitemOpen
  \bibfield  {author} {\bibinfo {author} {\bibfnamefont {P.}~\bibnamefont
  {Fendley}},\ }\bibfield  {title} {\bibinfo {title} {{Parafermionic edge zero
  modes in $\mathbb{Z}_n$-invariant spin chains}},\ }\href
  {https://doi.org/10.1088/1742-5468/2012/11/P11020} {\bibfield  {journal}
  {\bibinfo  {journal} {J. Stat. Mech.}\ }\textbf {\bibinfo {volume} {2012}},\
  \bibinfo {pages} {P11020} (\bibinfo {year} {2012})}\BibitemShut {NoStop}%
\bibitem [{\citenamefont {Fendley}(2014)}]{Fendley2014}%
  \BibitemOpen
  \bibfield  {author} {\bibinfo {author} {\bibfnamefont {P.}~\bibnamefont
  {Fendley}},\ }\bibfield  {title} {\bibinfo {title} {Free parafermions},\
  }\href {https://doi.org/10.1088/1751-8113/47/7/075001} {\bibfield  {journal}
  {\bibinfo  {journal} {J. Phys. A: Math. Theor.}\ }\textbf {\bibinfo {volume}
  {47}},\ \bibinfo {pages} {075001} (\bibinfo {year} {2014})}\BibitemShut
  {NoStop}%
\bibitem [{\citenamefont {Fendley}(2016)}]{Fendley2016}%
  \BibitemOpen
  \bibfield  {author} {\bibinfo {author} {\bibfnamefont {P.}~\bibnamefont
  {Fendley}},\ }\bibfield  {title} {\bibinfo {title} {Strong zero modes and
  eigenstate phase transitions in the xyz/interacting majorana chain},\ }\href
  {https://doi.org/10.1088/1751-8113/49/30/30LT01} {\bibfield  {journal}
  {\bibinfo  {journal} {J. Phys. A: Math. Theor.}\ }\textbf {\bibinfo {volume}
  {49}},\ \bibinfo {pages} {30LT01} (\bibinfo {year} {2016})}\BibitemShut
  {NoStop}%
\bibitem [{\citenamefont {Vasiloiu}\ \emph {et~al.}(2018)\citenamefont
  {Vasiloiu}, \citenamefont {Carollo},\ and\ \citenamefont
  {Garrahan}}]{Vasiloiu2018}%
  \BibitemOpen
  \bibfield  {author} {\bibinfo {author} {\bibfnamefont {L.~M.}\ \bibnamefont
  {Vasiloiu}}, \bibinfo {author} {\bibfnamefont {F.}~\bibnamefont {Carollo}},\
  and\ \bibinfo {author} {\bibfnamefont {J.~P.}\ \bibnamefont {Garrahan}},\
  }\bibfield  {title} {\bibinfo {title} {{Enhancing correlation times for edge
  spins through dissipation}},\ }\href
  {https://doi.org/10.1103/PhysRevB.98.094308} {\bibfield  {journal} {\bibinfo
  {journal} {Phys. Rev. B}\ }\textbf {\bibinfo {volume} {98}},\ \bibinfo
  {pages} {094308} (\bibinfo {year} {2018})}\BibitemShut {NoStop}%
\bibitem [{\citenamefont {Vasiloiu}\ \emph {et~al.}(2019)\citenamefont
  {Vasiloiu}, \citenamefont {Carollo}, \citenamefont {Marcuzzi},\ and\
  \citenamefont {Garrahan}}]{Vasiloiu2019}%
  \BibitemOpen
  \bibfield  {author} {\bibinfo {author} {\bibfnamefont {L.~M.}\ \bibnamefont
  {Vasiloiu}}, \bibinfo {author} {\bibfnamefont {F.}~\bibnamefont {Carollo}},
  \bibinfo {author} {\bibfnamefont {M.}~\bibnamefont {Marcuzzi}},\ and\
  \bibinfo {author} {\bibfnamefont {J.~P.}\ \bibnamefont {Garrahan}},\
  }\bibfield  {title} {\bibinfo {title} {{Strong zero modes in a class of
  generalized Ising spin ladders with plaquette interactions}},\ }\href
  {https://doi.org/10.1103/PhysRevB.100.024309} {\bibfield  {journal} {\bibinfo
   {journal} {Phys. Rev. B}\ }\textbf {\bibinfo {volume} {100}},\ \bibinfo
  {pages} {024309} (\bibinfo {year} {2019})}\BibitemShut {NoStop}%
\bibitem [{\citenamefont {Kemp}\ \emph {et~al.}(2020)\citenamefont {Kemp},
  \citenamefont {Yao},\ and\ \citenamefont {Laumann}}]{Kemp2020}%
  \BibitemOpen
  \bibfield  {author} {\bibinfo {author} {\bibfnamefont {J.}~\bibnamefont
  {Kemp}}, \bibinfo {author} {\bibfnamefont {N.~Y.}\ \bibnamefont {Yao}},\ and\
  \bibinfo {author} {\bibfnamefont {C.~R.}\ \bibnamefont {Laumann}},\
  }\bibfield  {title} {\bibinfo {title} {{Symmetry-Enhanced Boundary Qubits at
  Infinite Temperature}},\ }\href
  {https://doi.org/10.1103/PhysRevLett.125.200506} {\bibfield  {journal}
  {\bibinfo  {journal} {Phys. Rev. Lett.}\ }\textbf {\bibinfo {volume} {125}},\
  \bibinfo {pages} {200506} (\bibinfo {year} {2020})}\BibitemShut {NoStop}%
\bibitem [{\citenamefont {Kemp}\ \emph {et~al.}(2017)\citenamefont {Kemp},
  \citenamefont {Yao}, \citenamefont {Laumann},\ and\ \citenamefont
  {Fendley}}]{Kemp2017}%
  \BibitemOpen
  \bibfield  {author} {\bibinfo {author} {\bibfnamefont {J.}~\bibnamefont
  {Kemp}}, \bibinfo {author} {\bibfnamefont {N.~Y.}\ \bibnamefont {Yao}},
  \bibinfo {author} {\bibfnamefont {C.~R.}\ \bibnamefont {Laumann}},\ and\
  \bibinfo {author} {\bibfnamefont {P.}~\bibnamefont {Fendley}},\ }\bibfield
  {title} {\bibinfo {title} {Long coherence times for edge spins},\ }\bibfield
  {journal} {\bibinfo  {journal} {J. Stat. Mech.}\ }\href
  {https://doi.org/10.1088/1742-5468/aa73f0} {10.1088/1742-5468/aa73f0}
  (\bibinfo {year} {2017})\BibitemShut {NoStop}%
\bibitem [{\citenamefont {Chepiga}\ and\ \citenamefont
  {Laflorencie}(2023)}]{Chepiga2023}%
  \BibitemOpen
  \bibfield  {author} {\bibinfo {author} {\bibfnamefont {N.}~\bibnamefont
  {Chepiga}}\ and\ \bibinfo {author} {\bibfnamefont {N.}~\bibnamefont
  {Laflorencie}},\ }\bibfield  {title} {\bibinfo {title} {{Topological and
  quantum critical properties of the interacting Majorana chain model}},\
  }\href {https://doi.org/10.21468/SciPostPhys.14.6.152} {\bibfield  {journal}
  {\bibinfo  {journal} {SciPost Phys.}\ }\textbf {\bibinfo {volume} {14}},\
  \bibinfo {pages} {152} (\bibinfo {year} {2023})}\BibitemShut {NoStop}%
\bibitem [{\citenamefont {Laflorencie}(2023)}]{Laflorencie2023}%
  \BibitemOpen
  \bibfield  {author} {\bibinfo {author} {\bibfnamefont {N.}~\bibnamefont
  {Laflorencie}},\ }\href {https://arxiv.org/abs/2311.07571} {\bibinfo {title}
  {{Universal signatures of Majorana zero modes in critical Kitaev chains}}}
  (\bibinfo {year} {2023}),\ \Eprint {https://arxiv.org/abs/2311.07571}
  {arXiv:2311.07571} \BibitemShut {NoStop}%
\bibitem [{\citenamefont {Essler}\ \emph {et~al.}(2025)\citenamefont {Essler},
  \citenamefont {Fendley},\ and\ \citenamefont {Vernier}}]{Essler2025}%
  \BibitemOpen
  \bibfield  {author} {\bibinfo {author} {\bibfnamefont {F.~H.~L.}\
  \bibnamefont {Essler}}, \bibinfo {author} {\bibfnamefont {P.}~\bibnamefont
  {Fendley}},\ and\ \bibinfo {author} {\bibfnamefont {E.}~\bibnamefont
  {Vernier}},\ }\href {https://arxiv.org/abs/2512.07742} {\bibinfo {title}
  {{Strong zero modes in integrable spin-S chains}}} (\bibinfo {year} {2025}),\
  \Eprint {https://arxiv.org/abs/2512.07742} {arXiv:2512.07742} \BibitemShut
  {NoStop}%
\bibitem [{\citenamefont {Sreejith}\ \emph {et~al.}(2016)\citenamefont
  {Sreejith}, \citenamefont {Lazarides},\ and\ \citenamefont
  {Moessner}}]{Moessner2016}%
  \BibitemOpen
  \bibfield  {author} {\bibinfo {author} {\bibfnamefont {G.~J.}\ \bibnamefont
  {Sreejith}}, \bibinfo {author} {\bibfnamefont {A.}~\bibnamefont
  {Lazarides}},\ and\ \bibinfo {author} {\bibfnamefont {R.}~\bibnamefont
  {Moessner}},\ }\bibfield  {title} {\bibinfo {title} {{Parafermion chain with
  $2\ensuremath{\pi}/k$ Floquet edge modes}},\ }\href
  {https://doi.org/10.1103/PhysRevB.94.045127} {\bibfield  {journal} {\bibinfo
  {journal} {Phys. Rev. B}\ }\textbf {\bibinfo {volume} {94}},\ \bibinfo
  {pages} {045127} (\bibinfo {year} {2016})}\BibitemShut {NoStop}%
\bibitem [{\citenamefont {von Keyserlingk}\ and\ \citenamefont
  {Sondhi}(2016)}]{Keyserlingk2016}%
  \BibitemOpen
  \bibfield  {author} {\bibinfo {author} {\bibfnamefont {C.~W.}\ \bibnamefont
  {von Keyserlingk}}\ and\ \bibinfo {author} {\bibfnamefont {S.~L.}\
  \bibnamefont {Sondhi}},\ }\bibfield  {title} {\bibinfo {title} {{Phase
  structure of one-dimensional interacting Floquet systems. I. Abelian
  symmetry-protected topological phases}},\ }\href
  {https://doi.org/10.1103/PhysRevB.93.245145} {\bibfield  {journal} {\bibinfo
  {journal} {Phys. Rev. B}\ }\textbf {\bibinfo {volume} {93}},\ \bibinfo
  {pages} {245145} (\bibinfo {year} {2016})}\BibitemShut {NoStop}%
\bibitem [{\citenamefont {Yates}\ \emph {et~al.}(2020)\citenamefont {Yates},
  \citenamefont {Abanov},\ and\ \citenamefont {Mitra}}]{Yates2020}%
  \BibitemOpen
  \bibfield  {author} {\bibinfo {author} {\bibfnamefont {D.~J.}\ \bibnamefont
  {Yates}}, \bibinfo {author} {\bibfnamefont {A.~G.}\ \bibnamefont {Abanov}},\
  and\ \bibinfo {author} {\bibfnamefont {A.}~\bibnamefont {Mitra}},\ }\bibfield
   {title} {\bibinfo {title} {{Lifetime of Almost Strong Edge-Mode Operators in
  One-Dimensional, Interacting, Symmetry Protected Topological Phases}},\
  }\href {https://doi.org/10.1103/PhysRevLett.124.206803} {\bibfield  {journal}
  {\bibinfo  {journal} {Phys. Rev. Lett.}\ }\textbf {\bibinfo {volume} {124}},\
  \bibinfo {pages} {206803} (\bibinfo {year} {2020})}\BibitemShut {NoStop}%
\bibitem [{\citenamefont {Yates}\ \emph {et~al.}(2019)\citenamefont {Yates},
  \citenamefont {Essler},\ and\ \citenamefont {Mitra}}]{Yates2019}%
  \BibitemOpen
  \bibfield  {author} {\bibinfo {author} {\bibfnamefont {D.~J.}\ \bibnamefont
  {Yates}}, \bibinfo {author} {\bibfnamefont {F.~H.~L.}\ \bibnamefont
  {Essler}},\ and\ \bibinfo {author} {\bibfnamefont {A.}~\bibnamefont
  {Mitra}},\ }\bibfield  {title} {\bibinfo {title} {{Almost strong
  ($0,\ensuremath{\pi}$) edge modes in clean interacting one-dimensional
  Floquet systems}},\ }\href {https://doi.org/10.1103/PhysRevB.99.205419}
  {\bibfield  {journal} {\bibinfo  {journal} {Phys. Rev. B}\ }\textbf {\bibinfo
  {volume} {99}},\ \bibinfo {pages} {205419} (\bibinfo {year}
  {2019})}\BibitemShut {NoStop}%
\bibitem [{\citenamefont {Parker}\ \emph {et~al.}(2019)\citenamefont {Parker},
  \citenamefont {Vasseur},\ and\ \citenamefont {Scaffidi}}]{Parker2019}%
  \BibitemOpen
  \bibfield  {author} {\bibinfo {author} {\bibfnamefont {D.~E.}\ \bibnamefont
  {Parker}}, \bibinfo {author} {\bibfnamefont {R.}~\bibnamefont {Vasseur}},\
  and\ \bibinfo {author} {\bibfnamefont {T.}~\bibnamefont {Scaffidi}},\
  }\bibfield  {title} {\bibinfo {title} {{Topologically Protected Long Edge
  Coherence Times in Symmetry-Broken Phases}},\ }\href
  {https://doi.org/10.1103/PhysRevLett.122.240605} {\bibfield  {journal}
  {\bibinfo  {journal} {Phys. Rev. Lett.}\ }\textbf {\bibinfo {volume} {122}},\
  \bibinfo {pages} {240605} (\bibinfo {year} {2019})}\BibitemShut {NoStop}%
\bibitem [{\citenamefont {Yates}\ and\ \citenamefont
  {Mitra}(2021)}]{Yates2021}%
  \BibitemOpen
  \bibfield  {author} {\bibinfo {author} {\bibfnamefont {D.~J.}\ \bibnamefont
  {Yates}}\ and\ \bibinfo {author} {\bibfnamefont {A.}~\bibnamefont {Mitra}},\
  }\bibfield  {title} {\bibinfo {title} {{Strong and almost strong modes of
  Floquet spin chains in Krylov subspaces}},\ }\href
  {https://doi.org/10.1103/PhysRevB.104.195121} {\bibfield  {journal} {\bibinfo
   {journal} {Phys. Rev. B}\ }\textbf {\bibinfo {volume} {104}},\ \bibinfo
  {pages} {195121} (\bibinfo {year} {2021})}\BibitemShut {NoStop}%
\bibitem [{\citenamefont {{X. Mi \textit{et al.}}}(2022)}]{Mi2022}%
  \BibitemOpen
  \bibfield  {author} {\bibinfo {author} {\bibnamefont {{X. Mi \textit{et
  al.}}}},\ }\bibfield  {title} {\bibinfo {title} {Noise-resilient edge modes
  on a chain of superconducting qubits},\ }\href
  {https://doi.org/10.1126/science.abq5769} {\bibfield  {journal} {\bibinfo
  {journal} {Science}\ }\textbf {\bibinfo {volume} {378}},\ \bibinfo {pages}
  {785} (\bibinfo {year} {2022})}\BibitemShut {NoStop}%
\bibitem [{\citenamefont {Yeh}\ \emph {et~al.}(2023)\citenamefont {Yeh},
  \citenamefont {Rosch},\ and\ \citenamefont {Mitra}}]{Yeh2023}%
  \BibitemOpen
  \bibfield  {author} {\bibinfo {author} {\bibfnamefont {H.-C.}\ \bibnamefont
  {Yeh}}, \bibinfo {author} {\bibfnamefont {A.}~\bibnamefont {Rosch}},\ and\
  \bibinfo {author} {\bibfnamefont {A.}~\bibnamefont {Mitra}},\ }\bibfield
  {title} {\bibinfo {title} {{Decay rates of almost strong modes in Floquet
  spin chains beyond Fermi's Golden Rule}},\ }\href
  {https://doi.org/10.1103/PhysRevB.108.075112} {\bibfield  {journal} {\bibinfo
   {journal} {Phys. Rev. B}\ }\textbf {\bibinfo {volume} {108}},\ \bibinfo
  {pages} {075112} (\bibinfo {year} {2023})}\BibitemShut {NoStop}%
\bibitem [{\citenamefont {Yeh}\ \emph {et~al.}(2024)\citenamefont {Yeh},
  \citenamefont {Rosch},\ and\ \citenamefont {Mitra}}]{Yeh2024}%
  \BibitemOpen
  \bibfield  {author} {\bibinfo {author} {\bibfnamefont {H.-C.}\ \bibnamefont
  {Yeh}}, \bibinfo {author} {\bibfnamefont {A.}~\bibnamefont {Rosch}},\ and\
  \bibinfo {author} {\bibfnamefont {A.}~\bibnamefont {Mitra}},\ }\bibfield
  {title} {\bibinfo {title} {Floquet product mode},\ }\href
  {https://doi.org/10.1103/PhysRevB.110.075117} {\bibfield  {journal} {\bibinfo
   {journal} {Phys. Rev. B}\ }\textbf {\bibinfo {volume} {110}},\ \bibinfo
  {pages} {075117} (\bibinfo {year} {2024})}\BibitemShut {NoStop}%
\bibitem [{\citenamefont {Vernier}\ \emph {et~al.}(2024)\citenamefont
  {Vernier}, \citenamefont {Yeh}, \citenamefont {Piroli},\ and\ \citenamefont
  {Mitra}}]{Vernier2024}%
  \BibitemOpen
  \bibfield  {author} {\bibinfo {author} {\bibfnamefont {E.}~\bibnamefont
  {Vernier}}, \bibinfo {author} {\bibfnamefont {H.-C.}\ \bibnamefont {Yeh}},
  \bibinfo {author} {\bibfnamefont {L.}~\bibnamefont {Piroli}},\ and\ \bibinfo
  {author} {\bibfnamefont {A.}~\bibnamefont {Mitra}},\ }\bibfield  {title}
  {\bibinfo {title} {{Strong Zero Modes in Integrable Quantum Circuits}},\
  }\href {https://doi.org/10.1103/PhysRevLett.133.050606} {\bibfield  {journal}
  {\bibinfo  {journal} {Phys. Rev. Lett.}\ }\textbf {\bibinfo {volume} {133}},\
  \bibinfo {pages} {050606} (\bibinfo {year} {2024})}\BibitemShut {NoStop}%
\bibitem [{\citenamefont {Schmid}\ \emph {et~al.}(2024)\citenamefont {Schmid},
  \citenamefont {Penner}, \citenamefont {Yang}, \citenamefont {Glazman},\ and\
  \citenamefont {von Oppen}}]{Schmid2024}%
  \BibitemOpen
  \bibfield  {author} {\bibinfo {author} {\bibfnamefont {H.}~\bibnamefont
  {Schmid}}, \bibinfo {author} {\bibfnamefont {A.-G.}\ \bibnamefont {Penner}},
  \bibinfo {author} {\bibfnamefont {K.}~\bibnamefont {Yang}}, \bibinfo {author}
  {\bibfnamefont {L.}~\bibnamefont {Glazman}},\ and\ \bibinfo {author}
  {\bibfnamefont {F.}~\bibnamefont {von Oppen}},\ }\bibfield  {title} {\bibinfo
  {title} {{Robust Spectral $\ensuremath{\pi}$ Pairing in the Random-Field
  Floquet Quantum Ising Model}},\ }\href
  {https://doi.org/10.1103/PhysRevLett.132.210401} {\bibfield  {journal}
  {\bibinfo  {journal} {Phys. Rev. Lett.}\ }\textbf {\bibinfo {volume} {132}},\
  \bibinfo {pages} {210401} (\bibinfo {year} {2024})}\BibitemShut {NoStop}%
\bibitem [{\citenamefont {{F. Jin \textit{et al.}}}(2025)}]{Jin2025}%
  \BibitemOpen
  \bibfield  {author} {\bibinfo {author} {\bibnamefont {{F. Jin \textit{et
  al.}}}},\ }\bibfield  {title} {\bibinfo {title} {{Topological prethermal
  strong zero modes on superconducting processors}},\ }\href
  {https://doi.org/10.1038/s41586-025-09476-z} {\bibfield  {journal} {\bibinfo
  {journal} {Nature}\ }\textbf {\bibinfo {volume} {645}},\ \bibinfo {pages}
  {626} (\bibinfo {year} {2025})}\BibitemShut {NoStop}%
\bibitem [{\citenamefont {Friedman}\ \emph {et~al.}(2022)\citenamefont
  {Friedman}, \citenamefont {Ware}, \citenamefont {Vasseur},\ and\
  \citenamefont {Potter}}]{Friedman2022}%
  \BibitemOpen
  \bibfield  {author} {\bibinfo {author} {\bibfnamefont {A.~J.}\ \bibnamefont
  {Friedman}}, \bibinfo {author} {\bibfnamefont {B.}~\bibnamefont {Ware}},
  \bibinfo {author} {\bibfnamefont {R.}~\bibnamefont {Vasseur}},\ and\ \bibinfo
  {author} {\bibfnamefont {A.~C.}\ \bibnamefont {Potter}},\ }\bibfield  {title}
  {\bibinfo {title} {{Topological edge modes without symmetry in
  quasiperiodically driven spin chains}},\ }\href
  {https://doi.org/10.1103/PhysRevB.105.115117} {\bibfield  {journal} {\bibinfo
   {journal} {Phys. Rev. B}\ }\textbf {\bibinfo {volume} {105}},\ \bibinfo
  {pages} {115117} (\bibinfo {year} {2022})}\BibitemShut {NoStop}%
\bibitem [{\citenamefont {Schmid}\ \emph {et~al.}(2025)\citenamefont {Schmid},
  \citenamefont {Peng}, \citenamefont {Refael},\ and\ \citenamefont {von
  Oppen}}]{Schmid2024a}%
  \BibitemOpen
  \bibfield  {author} {\bibinfo {author} {\bibfnamefont {H.}~\bibnamefont
  {Schmid}}, \bibinfo {author} {\bibfnamefont {Y.}~\bibnamefont {Peng}},
  \bibinfo {author} {\bibfnamefont {G.}~\bibnamefont {Refael}},\ and\ \bibinfo
  {author} {\bibfnamefont {F.}~\bibnamefont {von Oppen}},\ }\bibfield  {title}
  {\bibinfo {title} {{Self-Similar Phase Diagram of the Fibonacci-Driven
  Quantum Ising Model}},\ }\href {https://doi.org/10.1103/hn66-j8pt} {\bibfield
   {journal} {\bibinfo  {journal} {Phys. Rev. Lett.}\ }\textbf {\bibinfo
  {volume} {134}},\ \bibinfo {pages} {240404} (\bibinfo {year}
  {2025})}\BibitemShut {NoStop}%
\bibitem [{\citenamefont {Klobas}\ \emph {et~al.}(2023)\citenamefont {Klobas},
  \citenamefont {Fendley},\ and\ \citenamefont {Garrahan}}]{Klobas2023}%
  \BibitemOpen
  \bibfield  {author} {\bibinfo {author} {\bibfnamefont {K.}~\bibnamefont
  {Klobas}}, \bibinfo {author} {\bibfnamefont {P.}~\bibnamefont {Fendley}},\
  and\ \bibinfo {author} {\bibfnamefont {J.~P.}\ \bibnamefont {Garrahan}},\
  }\bibfield  {title} {\bibinfo {title} {{Stochastic strong zero modes and
  their dynamical manifestations}},\ }\href
  {https://doi.org/10.1103/PhysRevE.107.L042104} {\bibfield  {journal}
  {\bibinfo  {journal} {Phys. Rev. E}\ }\textbf {\bibinfo {volume} {107}},\
  \bibinfo {pages} {L042104} (\bibinfo {year} {2023})}\BibitemShut {NoStop}%
\bibitem [{\citenamefont {Kitaev}(2001)}]{Kitaev2001}%
  \BibitemOpen
  \bibfield  {author} {\bibinfo {author} {\bibfnamefont {A.~Y.}\ \bibnamefont
  {Kitaev}},\ }\bibfield  {title} {\bibinfo {title} {{Unpaired Majorana
  fermions in quantum wires}},\ }\href
  {https://doi.org/10.1070/1063-7869/44/10s/s29} {\bibfield  {journal}
  {\bibinfo  {journal} {Physics-Uspekhi}\ }\textbf {\bibinfo {volume} {44}},\
  \bibinfo {pages} {131} (\bibinfo {year} {2001})}\BibitemShut {NoStop}%
\bibitem [{\citenamefont {Jiang}\ \emph {et~al.}(2011)\citenamefont {Jiang},
  \citenamefont {Kitagawa}, \citenamefont {Alicea}, \citenamefont {Akhmerov},
  \citenamefont {Pekker}, \citenamefont {Refael}, \citenamefont {Cirac},
  \citenamefont {Demler}, \citenamefont {Lukin},\ and\ \citenamefont
  {Zoller}}]{Jiang2011}%
  \BibitemOpen
  \bibfield  {author} {\bibinfo {author} {\bibfnamefont {L.}~\bibnamefont
  {Jiang}}, \bibinfo {author} {\bibfnamefont {T.}~\bibnamefont {Kitagawa}},
  \bibinfo {author} {\bibfnamefont {J.}~\bibnamefont {Alicea}}, \bibinfo
  {author} {\bibfnamefont {A.~R.}\ \bibnamefont {Akhmerov}}, \bibinfo {author}
  {\bibfnamefont {D.}~\bibnamefont {Pekker}}, \bibinfo {author} {\bibfnamefont
  {G.}~\bibnamefont {Refael}}, \bibinfo {author} {\bibfnamefont {J.~I.}\
  \bibnamefont {Cirac}}, \bibinfo {author} {\bibfnamefont {E.}~\bibnamefont
  {Demler}}, \bibinfo {author} {\bibfnamefont {M.~D.}\ \bibnamefont {Lukin}},\
  and\ \bibinfo {author} {\bibfnamefont {P.}~\bibnamefont {Zoller}},\
  }\bibfield  {title} {\bibinfo {title} {Majorana fermions in equilibrium and
  in driven cold-atom quantum wires},\ }\href
  {https://doi.org/10.1103/PhysRevLett.106.220402} {\bibfield  {journal}
  {\bibinfo  {journal} {Phys. Rev. Lett.}\ }\textbf {\bibinfo {volume} {106}},\
  \bibinfo {pages} {220402} (\bibinfo {year} {2011})}\BibitemShut {NoStop}%
\bibitem [{\citenamefont {Thakurathi}\ \emph {et~al.}(2013)\citenamefont
  {Thakurathi}, \citenamefont {Patel}, \citenamefont {Sen},\ and\ \citenamefont
  {Dutta}}]{Dutta2013}%
  \BibitemOpen
  \bibfield  {author} {\bibinfo {author} {\bibfnamefont {M.}~\bibnamefont
  {Thakurathi}}, \bibinfo {author} {\bibfnamefont {A.~A.}\ \bibnamefont
  {Patel}}, \bibinfo {author} {\bibfnamefont {D.}~\bibnamefont {Sen}},\ and\
  \bibinfo {author} {\bibfnamefont {A.}~\bibnamefont {Dutta}},\ }\bibfield
  {title} {\bibinfo {title} {{Floquet generation of Majorana end modes and
  topological invariants}},\ }\href
  {https://doi.org/10.1103/PhysRevB.88.155133} {\bibfield  {journal} {\bibinfo
  {journal} {Phys. Rev. B}\ }\textbf {\bibinfo {volume} {88}},\ \bibinfo
  {pages} {155133} (\bibinfo {year} {2013})}\BibitemShut {NoStop}%
\bibitem [{\citenamefont {Bauer}\ \emph {et~al.}(2019)\citenamefont {Bauer},
  \citenamefont {Pereg-Barnea}, \citenamefont {Karzig}, \citenamefont {Rieder},
  \citenamefont {Refael}, \citenamefont {Berg},\ and\ \citenamefont
  {Oreg}}]{Bauer2019}%
  \BibitemOpen
  \bibfield  {author} {\bibinfo {author} {\bibfnamefont {B.}~\bibnamefont
  {Bauer}}, \bibinfo {author} {\bibfnamefont {T.}~\bibnamefont {Pereg-Barnea}},
  \bibinfo {author} {\bibfnamefont {T.}~\bibnamefont {Karzig}}, \bibinfo
  {author} {\bibfnamefont {M.-T.}\ \bibnamefont {Rieder}}, \bibinfo {author}
  {\bibfnamefont {G.}~\bibnamefont {Refael}}, \bibinfo {author} {\bibfnamefont
  {E.}~\bibnamefont {Berg}},\ and\ \bibinfo {author} {\bibfnamefont
  {Y.}~\bibnamefont {Oreg}},\ }\bibfield  {title} {\bibinfo {title}
  {{Topologically protected braiding in a single wire using Floquet Majorana
  modes}},\ }\href {https://doi.org/10.1103/PhysRevB.100.041102} {\bibfield
  {journal} {\bibinfo  {journal} {Phys. Rev. B}\ }\textbf {\bibinfo {volume}
  {100}},\ \bibinfo {pages} {041102} (\bibinfo {year} {2019})}\BibitemShut
  {NoStop}%
\bibitem [{\citenamefont {Huse}\ \emph {et~al.}(2013)\citenamefont {Huse},
  \citenamefont {Nandkishore}, \citenamefont {Oganesyan}, \citenamefont {Pal},\
  and\ \citenamefont {Sondhi}}]{Huse2013}%
  \BibitemOpen
  \bibfield  {author} {\bibinfo {author} {\bibfnamefont {D.~A.}\ \bibnamefont
  {Huse}}, \bibinfo {author} {\bibfnamefont {R.}~\bibnamefont {Nandkishore}},
  \bibinfo {author} {\bibfnamefont {V.}~\bibnamefont {Oganesyan}}, \bibinfo
  {author} {\bibfnamefont {A.}~\bibnamefont {Pal}},\ and\ \bibinfo {author}
  {\bibfnamefont {S.~L.}\ \bibnamefont {Sondhi}},\ }\bibfield  {title}
  {\bibinfo {title} {Localization-protected quantum order},\ }\href
  {https://doi.org/10.1103/PhysRevB.88.014206} {\bibfield  {journal} {\bibinfo
  {journal} {Phys. Rev. B}\ }\textbf {\bibinfo {volume} {88}},\ \bibinfo
  {pages} {014206} (\bibinfo {year} {2013})}\BibitemShut {NoStop}%
\bibitem [{\citenamefont {Pekker}\ \emph {et~al.}(2014)\citenamefont {Pekker},
  \citenamefont {Refael}, \citenamefont {Altman}, \citenamefont {Demler},\ and\
  \citenamefont {Oganesyan}}]{Pekker2014}%
  \BibitemOpen
  \bibfield  {author} {\bibinfo {author} {\bibfnamefont {D.}~\bibnamefont
  {Pekker}}, \bibinfo {author} {\bibfnamefont {G.}~\bibnamefont {Refael}},
  \bibinfo {author} {\bibfnamefont {E.}~\bibnamefont {Altman}}, \bibinfo
  {author} {\bibfnamefont {E.}~\bibnamefont {Demler}},\ and\ \bibinfo {author}
  {\bibfnamefont {V.}~\bibnamefont {Oganesyan}},\ }\bibfield  {title} {\bibinfo
  {title} {{Hilbert-Glass Transition: New Universality of Temperature-Tuned
  Many-Body Dynamical Quantum Criticality}},\ }\href
  {https://doi.org/10.1103/PhysRevX.4.011052} {\bibfield  {journal} {\bibinfo
  {journal} {Phys. Rev. X}\ }\textbf {\bibinfo {volume} {4}},\ \bibinfo {pages}
  {011052} (\bibinfo {year} {2014})}\BibitemShut {NoStop}%
\bibitem [{\citenamefont {Khemani}\ \emph {et~al.}(2016)\citenamefont
  {Khemani}, \citenamefont {Lazarides}, \citenamefont {Moessner},\ and\
  \citenamefont {Sondhi}}]{Khemani2016}%
  \BibitemOpen
  \bibfield  {author} {\bibinfo {author} {\bibfnamefont {V.}~\bibnamefont
  {Khemani}}, \bibinfo {author} {\bibfnamefont {A.}~\bibnamefont {Lazarides}},
  \bibinfo {author} {\bibfnamefont {R.}~\bibnamefont {Moessner}},\ and\
  \bibinfo {author} {\bibfnamefont {S.~L.}\ \bibnamefont {Sondhi}},\ }\bibfield
   {title} {\bibinfo {title} {{Phase Structure of Driven Quantum Systems}},\
  }\href {https://doi.org/10.1103/PhysRevLett.116.250401} {\bibfield  {journal}
  {\bibinfo  {journal} {Phys. Rev. Lett.}\ }\textbf {\bibinfo {volume} {116}},\
  \bibinfo {pages} {250401} (\bibinfo {year} {2016})}\BibitemShut {NoStop}%
\bibitem [{\citenamefont {Mi}\ \emph {et~al.}(2022)\citenamefont {Mi},
  \citenamefont {Ippoliti}, \citenamefont {Quintana} \emph
  {et~al.}}]{mi2022time}%
  \BibitemOpen
  \bibfield  {author} {\bibinfo {author} {\bibfnamefont {X.}~\bibnamefont
  {Mi}}, \bibinfo {author} {\bibfnamefont {M.}~\bibnamefont {Ippoliti}},
  \bibinfo {author} {\bibfnamefont {C.}~\bibnamefont {Quintana}}, \emph
  {et~al.},\ }\bibfield  {title} {\bibinfo {title} {{Time-crystalline
  eigenstate order on a quantum processor}},\ }\href
  {https://doi.org/10.1038/s41586-021-04257-w} {\bibfield  {journal} {\bibinfo
  {journal} {Nature}\ }\textbf {\bibinfo {volume} {601}},\ \bibinfo {pages}
  {531} (\bibinfo {year} {2022})}\BibitemShut {NoStop}%
\bibitem [{\citenamefont {Lerose}\ \emph {et~al.}(2021)\citenamefont {Lerose},
  \citenamefont {Sonner},\ and\ \citenamefont {Abanin}}]{Lerose2021}%
  \BibitemOpen
  \bibfield  {author} {\bibinfo {author} {\bibfnamefont {A.}~\bibnamefont
  {Lerose}}, \bibinfo {author} {\bibfnamefont {M.}~\bibnamefont {Sonner}},\
  and\ \bibinfo {author} {\bibfnamefont {D.~A.}\ \bibnamefont {Abanin}},\
  }\bibfield  {title} {\bibinfo {title} {Scaling of temporal entanglement in
  proximity to integrability},\ }\href
  {https://doi.org/10.1103/PhysRevB.104.035137} {\bibfield  {journal} {\bibinfo
   {journal} {Phys. Rev. B}\ }\textbf {\bibinfo {volume} {104}},\ \bibinfo
  {pages} {035137} (\bibinfo {year} {2021})}\BibitemShut {NoStop}%
\bibitem [{\citenamefont {Penner}\ \emph {et~al.}(2025)\citenamefont {Penner},
  \citenamefont {Schmid}, \citenamefont {Glazman},\ and\ \citenamefont {von
  Oppen}}]{Penner2025}%
  \BibitemOpen
  \bibfield  {author} {\bibinfo {author} {\bibfnamefont {A.-G.}\ \bibnamefont
  {Penner}}, \bibinfo {author} {\bibfnamefont {H.}~\bibnamefont {Schmid}},
  \bibinfo {author} {\bibfnamefont {L.~I.}\ \bibnamefont {Glazman}},\ and\
  \bibinfo {author} {\bibfnamefont {F.}~\bibnamefont {von Oppen}},\ }\bibfield
  {title} {\bibinfo {title} {{Subharmonic spin correlations and spectral
  pairing in Floquet time crystals}},\ }\href
  {https://doi.org/10.1103/PhysRevB.111.184308} {\bibfield  {journal} {\bibinfo
   {journal} {Phys. Rev. B}\ }\textbf {\bibinfo {volume} {111}},\ \bibinfo
  {pages} {184308} (\bibinfo {year} {2025})}\BibitemShut {NoStop}%
\bibitem [{\citenamefont {{Bouchaud}}\ and\ \citenamefont
  {{Georges}}(1990)}]{Bouchaud1990}%
  \BibitemOpen
  \bibfield  {author} {\bibinfo {author} {\bibfnamefont {J.-P.}\ \bibnamefont
  {{Bouchaud}}}\ and\ \bibinfo {author} {\bibfnamefont {A.}~\bibnamefont
  {{Georges}}},\ }\bibfield  {title} {\bibinfo {title} {{Anomalous diffusion in
  disordered media: Statistical mechanisms, models and physical
  applications}},\ }\href {https://doi.org/10.1016/0370-1573(90)90099-N}
  {\bibfield  {journal} {\bibinfo  {journal} {Phys. Rep.}\ }\textbf {\bibinfo
  {volume} {195}},\ \bibinfo {pages} {127} (\bibinfo {year}
  {1990})}\BibitemShut {NoStop}%
\bibitem [{\citenamefont {Penner}\ \emph {et~al.}(2021)\citenamefont {Penner},
  \citenamefont {von Oppen}, \citenamefont {Zar\'and},\ and\ \citenamefont
  {Zirnbauer}}]{Penner2021}%
  \BibitemOpen
  \bibfield  {author} {\bibinfo {author} {\bibfnamefont {A.-G.}\ \bibnamefont
  {Penner}}, \bibinfo {author} {\bibfnamefont {F.}~\bibnamefont {von Oppen}},
  \bibinfo {author} {\bibfnamefont {G.}~\bibnamefont {Zar\'and}},\ and\
  \bibinfo {author} {\bibfnamefont {M.~R.}\ \bibnamefont {Zirnbauer}},\
  }\bibfield  {title} {\bibinfo {title} {Hilbert space geometry of random
  matrix eigenstates},\ }\href {https://doi.org/10.1103/PhysRevLett.126.200604}
  {\bibfield  {journal} {\bibinfo  {journal} {Phys. Rev. Lett.}\ }\textbf
  {\bibinfo {volume} {126}},\ \bibinfo {pages} {200604} (\bibinfo {year}
  {2021})}\BibitemShut {NoStop}%
\bibitem [{\citenamefont {Jermyn}\ \emph {et~al.}(2014)\citenamefont {Jermyn},
  \citenamefont {Mong}, \citenamefont {Alicea},\ and\ \citenamefont
  {Fendley}}]{Jermyn2014}%
  \BibitemOpen
  \bibfield  {author} {\bibinfo {author} {\bibfnamefont {A.~S.}\ \bibnamefont
  {Jermyn}}, \bibinfo {author} {\bibfnamefont {R.~S.~K.}\ \bibnamefont {Mong}},
  \bibinfo {author} {\bibfnamefont {J.}~\bibnamefont {Alicea}},\ and\ \bibinfo
  {author} {\bibfnamefont {P.}~\bibnamefont {Fendley}},\ }\bibfield  {title}
  {\bibinfo {title} {{Stability of zero modes in parafermion chains}},\ }\href
  {https://doi.org/10.1103/PhysRevB.90.165106} {\bibfield  {journal} {\bibinfo
  {journal} {Phys. Rev. B}\ }\textbf {\bibinfo {volume} {90}},\ \bibinfo
  {pages} {165106} (\bibinfo {year} {2014})}\BibitemShut {NoStop}%
\bibitem [{\citenamefont {Alicea}\ and\ \citenamefont
  {Fendley}(2016)}]{Alicea2016}%
  \BibitemOpen
  \bibfield  {author} {\bibinfo {author} {\bibfnamefont {J.}~\bibnamefont
  {Alicea}}\ and\ \bibinfo {author} {\bibfnamefont {P.}~\bibnamefont
  {Fendley}},\ }\bibfield  {title} {\bibinfo {title} {{Topological phases with
  parafermions: theory and blueprints}},\ }\href
  {https://doi.org/10.1146/annurev-conmatphys-031115-011336} {\bibfield
  {journal} {\bibinfo  {journal} {Annu. Rev. Condens. Matter Phys.}\ }\textbf
  {\bibinfo {volume} {7}},\ \bibinfo {pages} {119} (\bibinfo {year}
  {2016})}\BibitemShut {NoStop}%
\bibitem [{\citenamefont {Surace}\ \emph {et~al.}(2019)\citenamefont {Surace},
  \citenamefont {Russomanno}, \citenamefont {Dalmonte}, \citenamefont {Silva},
  \citenamefont {Fazio},\ and\ \citenamefont {Iemini}}]{Surace2019}%
  \BibitemOpen
  \bibfield  {author} {\bibinfo {author} {\bibfnamefont {F.~M.}\ \bibnamefont
  {Surace}}, \bibinfo {author} {\bibfnamefont {A.}~\bibnamefont {Russomanno}},
  \bibinfo {author} {\bibfnamefont {M.}~\bibnamefont {Dalmonte}}, \bibinfo
  {author} {\bibfnamefont {A.}~\bibnamefont {Silva}}, \bibinfo {author}
  {\bibfnamefont {R.}~\bibnamefont {Fazio}},\ and\ \bibinfo {author}
  {\bibfnamefont {F.}~\bibnamefont {Iemini}},\ }\bibfield  {title} {\bibinfo
  {title} {Floquet time crystals in clock models},\ }\href
  {https://doi.org/10.1103/PhysRevB.99.104303} {\bibfield  {journal} {\bibinfo
  {journal} {Phys. Rev. B}\ }\textbf {\bibinfo {volume} {99}},\ \bibinfo
  {pages} {104303} (\bibinfo {year} {2019})}\BibitemShut {NoStop}%
\bibitem [{\citenamefont {Verdeny}\ \emph {et~al.}(2016)\citenamefont
  {Verdeny}, \citenamefont {Puig},\ and\ \citenamefont
  {Mintert}}]{Verdeny2016}%
  \BibitemOpen
  \bibfield  {author} {\bibinfo {author} {\bibfnamefont {A.}~\bibnamefont
  {Verdeny}}, \bibinfo {author} {\bibfnamefont {J.}~\bibnamefont {Puig}},\ and\
  \bibinfo {author} {\bibfnamefont {F.}~\bibnamefont {Mintert}},\ }\bibfield
  {title} {\bibinfo {title} {{Quasi-Periodically Driven Quantum Systems}},\
  }\href {https://doi.org/doi:10.1515/zna-2016-0079} {\bibfield  {journal}
  {\bibinfo  {journal} {Z.\ Naturforsch.\ A}\ }\textbf {\bibinfo {volume}
  {71}},\ \bibinfo {pages} {897} (\bibinfo {year} {2016})}\BibitemShut
  {NoStop}%
\bibitem [{\citenamefont {Dumitrescu}\ \emph {et~al.}(2018)\citenamefont
  {Dumitrescu}, \citenamefont {Vasseur},\ and\ \citenamefont
  {Potter}}]{Dumitrescu2018}%
  \BibitemOpen
  \bibfield  {author} {\bibinfo {author} {\bibfnamefont {P.~T.}\ \bibnamefont
  {Dumitrescu}}, \bibinfo {author} {\bibfnamefont {R.}~\bibnamefont
  {Vasseur}},\ and\ \bibinfo {author} {\bibfnamefont {A.~C.}\ \bibnamefont
  {Potter}},\ }\bibfield  {title} {\bibinfo {title} {{Logarithmically Slow
  Relaxation in Quasiperiodically Driven Random Spin Chains}},\ }\href
  {https://doi.org/10.1103/PhysRevLett.120.070602} {\bibfield  {journal}
  {\bibinfo  {journal} {Phys. Rev. Lett.}\ }\textbf {\bibinfo {volume} {120}},\
  \bibinfo {pages} {070602} (\bibinfo {year} {2018})}\BibitemShut {NoStop}%
\bibitem [{\citenamefont {Peng}\ and\ \citenamefont {Refael}(2018)}]{Peng2018}%
  \BibitemOpen
  \bibfield  {author} {\bibinfo {author} {\bibfnamefont {Y.}~\bibnamefont
  {Peng}}\ and\ \bibinfo {author} {\bibfnamefont {G.}~\bibnamefont {Refael}},\
  }\bibfield  {title} {\bibinfo {title} {{Time-quasiperiodic topological
  superconductors with Majorana multiplexing}},\ }\href
  {https://doi.org/10.1103/PhysRevB.98.220509} {\bibfield  {journal} {\bibinfo
  {journal} {Phys. Rev. B}\ }\textbf {\bibinfo {volume} {98}},\ \bibinfo
  {pages} {220509} (\bibinfo {year} {2018})}\BibitemShut {NoStop}%
\bibitem [{\citenamefont {Crowley}\ \emph {et~al.}(2019)\citenamefont
  {Crowley}, \citenamefont {Martin},\ and\ \citenamefont
  {Chandran}}]{Crowley2019}%
  \BibitemOpen
  \bibfield  {author} {\bibinfo {author} {\bibfnamefont {P.~J.~D.}\
  \bibnamefont {Crowley}}, \bibinfo {author} {\bibfnamefont {I.}~\bibnamefont
  {Martin}},\ and\ \bibinfo {author} {\bibfnamefont {A.}~\bibnamefont
  {Chandran}},\ }\bibfield  {title} {\bibinfo {title} {Topological
  classification of quasiperiodically driven quantum systems},\ }\href
  {https://doi.org/10.1103/PhysRevB.99.064306} {\bibfield  {journal} {\bibinfo
  {journal} {Phys. Rev. B}\ }\textbf {\bibinfo {volume} {99}},\ \bibinfo
  {pages} {064306} (\bibinfo {year} {2019})}\BibitemShut {NoStop}%
\bibitem [{\citenamefont {Maity}\ \emph {et~al.}(2019)\citenamefont {Maity},
  \citenamefont {Bhattacharya}, \citenamefont {Dutta},\ and\ \citenamefont
  {Sen}}]{Maity2019}%
  \BibitemOpen
  \bibfield  {author} {\bibinfo {author} {\bibfnamefont {S.}~\bibnamefont
  {Maity}}, \bibinfo {author} {\bibfnamefont {U.}~\bibnamefont {Bhattacharya}},
  \bibinfo {author} {\bibfnamefont {A.}~\bibnamefont {Dutta}},\ and\ \bibinfo
  {author} {\bibfnamefont {D.}~\bibnamefont {Sen}},\ }\bibfield  {title}
  {\bibinfo {title} {{Fibonacci steady states in a driven integrable quantum
  system}},\ }\href {https://doi.org/10.1103/PhysRevB.99.020306} {\bibfield
  {journal} {\bibinfo  {journal} {Phys. Rev. B}\ }\textbf {\bibinfo {volume}
  {99}},\ \bibinfo {pages} {020306} (\bibinfo {year} {2019})}\BibitemShut
  {NoStop}%
\bibitem [{\citenamefont {Lapierre}\ \emph {et~al.}(2020)\citenamefont
  {Lapierre}, \citenamefont {Choo}, \citenamefont {Tiwari}, \citenamefont
  {Tauber}, \citenamefont {Neupert},\ and\ \citenamefont
  {Chitra}}]{Lapierre2020}%
  \BibitemOpen
  \bibfield  {author} {\bibinfo {author} {\bibfnamefont {B.}~\bibnamefont
  {Lapierre}}, \bibinfo {author} {\bibfnamefont {K.}~\bibnamefont {Choo}},
  \bibinfo {author} {\bibfnamefont {A.}~\bibnamefont {Tiwari}}, \bibinfo
  {author} {\bibfnamefont {C.}~\bibnamefont {Tauber}}, \bibinfo {author}
  {\bibfnamefont {T.}~\bibnamefont {Neupert}},\ and\ \bibinfo {author}
  {\bibfnamefont {R.}~\bibnamefont {Chitra}},\ }\bibfield  {title} {\bibinfo
  {title} {Fine structure of heating in a quasiperiodically driven critical
  quantum system},\ }\href {https://doi.org/10.1103/PhysRevResearch.2.033461}
  {\bibfield  {journal} {\bibinfo  {journal} {Phys. Rev. Res.}\ }\textbf
  {\bibinfo {volume} {2}},\ \bibinfo {pages} {033461} (\bibinfo {year}
  {2020})}\BibitemShut {NoStop}%
\bibitem [{\citenamefont {Long}\ \emph {et~al.}(2021)\citenamefont {Long},
  \citenamefont {Crowley},\ and\ \citenamefont {Chandran}}]{Long2021}%
  \BibitemOpen
  \bibfield  {author} {\bibinfo {author} {\bibfnamefont {D.~M.}\ \bibnamefont
  {Long}}, \bibinfo {author} {\bibfnamefont {P.~J.~D.}\ \bibnamefont
  {Crowley}},\ and\ \bibinfo {author} {\bibfnamefont {A.}~\bibnamefont
  {Chandran}},\ }\bibfield  {title} {\bibinfo {title} {{Nonadiabatic
  Topological Energy Pumps with Quasiperiodic Driving}},\ }\href
  {https://doi.org/10.1103/PhysRevLett.126.106805} {\bibfield  {journal}
  {\bibinfo  {journal} {Phys. Rev. Lett.}\ }\textbf {\bibinfo {volume} {126}},\
  \bibinfo {pages} {106805} (\bibinfo {year} {2021})}\BibitemShut {NoStop}%
\bibitem [{\citenamefont {Bhattacharjee}\ \emph {et~al.}(2022)\citenamefont
  {Bhattacharjee}, \citenamefont {Bandyopadhyay},\ and\ \citenamefont
  {Dutta}}]{Bhattacharjee2022}%
  \BibitemOpen
  \bibfield  {author} {\bibinfo {author} {\bibfnamefont {S.}~\bibnamefont
  {Bhattacharjee}}, \bibinfo {author} {\bibfnamefont {S.}~\bibnamefont
  {Bandyopadhyay}},\ and\ \bibinfo {author} {\bibfnamefont {A.}~\bibnamefont
  {Dutta}},\ }\bibfield  {title} {\bibinfo {title} {{Quasilocalization dynamics
  in a Fibonacci quantum rotor}},\ }\href
  {https://doi.org/10.1103/PhysRevA.106.022206} {\bibfield  {journal} {\bibinfo
   {journal} {Phys. Rev. A}\ }\textbf {\bibinfo {volume} {106}},\ \bibinfo
  {pages} {022206} (\bibinfo {year} {2022})}\BibitemShut {NoStop}%
\bibitem [{\citenamefont {Zhao}\ \emph {et~al.}(2021)\citenamefont {Zhao},
  \citenamefont {Mintert}, \citenamefont {Moessner},\ and\ \citenamefont
  {Knolle}}]{Zhao2021}%
  \BibitemOpen
  \bibfield  {author} {\bibinfo {author} {\bibfnamefont {H.}~\bibnamefont
  {Zhao}}, \bibinfo {author} {\bibfnamefont {F.}~\bibnamefont {Mintert}},
  \bibinfo {author} {\bibfnamefont {R.}~\bibnamefont {Moessner}},\ and\
  \bibinfo {author} {\bibfnamefont {J.}~\bibnamefont {Knolle}},\ }\bibfield
  {title} {\bibinfo {title} {{Random Multipolar Driving: Tunably Slow Heating
  through Spectral Engineering}},\ }\href
  {https://doi.org/10.1103/PhysRevLett.126.040601} {\bibfield  {journal}
  {\bibinfo  {journal} {Phys. Rev. Lett.}\ }\textbf {\bibinfo {volume} {126}},\
  \bibinfo {pages} {040601} (\bibinfo {year} {2021})}\BibitemShut {NoStop}%
\bibitem [{\citenamefont {Zhao}\ \emph {et~al.}(2022)\citenamefont {Zhao},
  \citenamefont {Rudner}, \citenamefont {Moessner},\ and\ \citenamefont
  {Knolle}}]{Zhao2022}%
  \BibitemOpen
  \bibfield  {author} {\bibinfo {author} {\bibfnamefont {H.}~\bibnamefont
  {Zhao}}, \bibinfo {author} {\bibfnamefont {M.~S.}\ \bibnamefont {Rudner}},
  \bibinfo {author} {\bibfnamefont {R.}~\bibnamefont {Moessner}},\ and\
  \bibinfo {author} {\bibfnamefont {J.}~\bibnamefont {Knolle}},\ }\bibfield
  {title} {\bibinfo {title} {Anomalous random multipolar driven insulators},\
  }\href {https://doi.org/10.1103/PhysRevB.105.245119} {\bibfield  {journal}
  {\bibinfo  {journal} {Phys. Rev. B}\ }\textbf {\bibinfo {volume} {105}},\
  \bibinfo {pages} {245119} (\bibinfo {year} {2022})}\BibitemShut {NoStop}%
\bibitem [{\citenamefont {Zhao}\ \emph {et~al.}(2023)\citenamefont {Zhao},
  \citenamefont {Knolle},\ and\ \citenamefont {Moessner}}]{Zhao2023}%
  \BibitemOpen
  \bibfield  {author} {\bibinfo {author} {\bibfnamefont {H.}~\bibnamefont
  {Zhao}}, \bibinfo {author} {\bibfnamefont {J.}~\bibnamefont {Knolle}},\ and\
  \bibinfo {author} {\bibfnamefont {R.}~\bibnamefont {Moessner}},\ }\bibfield
  {title} {\bibinfo {title} {Temporal disorder in spatiotemporal order},\
  }\href {https://doi.org/10.1103/PhysRevB.108.L100203} {\bibfield  {journal}
  {\bibinfo  {journal} {Phys. Rev. B}\ }\textbf {\bibinfo {volume} {108}},\
  \bibinfo {pages} {L100203} (\bibinfo {year} {2023})}\BibitemShut {NoStop}%
\bibitem [{\citenamefont {{Z.-H. Liu \textit{et al.}}}(2026)}]{Liu2025}%
  \BibitemOpen
  \bibfield  {author} {\bibinfo {author} {\bibnamefont {{Z.-H. Liu \textit{et
  al.}}}},\ }\bibfield  {title} {\bibinfo {title} {Prethermalization by random
  multipolar driving on a 78-qubit processor},\ }\href
  {https://doi.org/10.1038/s41586-025-09977-x} {\bibfield  {journal} {\bibinfo
  {journal} {Nature}\ }\textbf {\bibinfo {volume} {650}},\ \bibinfo {pages}
  {79} (\bibinfo {year} {2026})}\BibitemShut {NoStop}%
\bibitem [{\citenamefont {Wen}\ \emph {et~al.}(2021)\citenamefont {Wen},
  \citenamefont {Fan}, \citenamefont {Vishwanath},\ and\ \citenamefont
  {Gu}}]{Wen2021}%
  \BibitemOpen
  \bibfield  {author} {\bibinfo {author} {\bibfnamefont {X.}~\bibnamefont
  {Wen}}, \bibinfo {author} {\bibfnamefont {R.}~\bibnamefont {Fan}}, \bibinfo
  {author} {\bibfnamefont {A.}~\bibnamefont {Vishwanath}},\ and\ \bibinfo
  {author} {\bibfnamefont {Y.}~\bibnamefont {Gu}},\ }\bibfield  {title}
  {\bibinfo {title} {Periodically, quasiperiodically, and randomly driven
  conformal field theories},\ }\href
  {https://doi.org/10.1103/PhysRevResearch.3.023044} {\bibfield  {journal}
  {\bibinfo  {journal} {Phys. Rev. Res.}\ }\textbf {\bibinfo {volume} {3}},\
  \bibinfo {pages} {023044} (\bibinfo {year} {2021})}\BibitemShut {NoStop}%
\bibitem [{\citenamefont {Lapierre}\ \emph
  {et~al.}(2025{\natexlab{a}})\citenamefont {Lapierre}, \citenamefont {Mo},\
  and\ \citenamefont {Ryu}}]{Lapierre2025a}%
  \BibitemOpen
  \bibfield  {author} {\bibinfo {author} {\bibfnamefont {B.}~\bibnamefont
  {Lapierre}}, \bibinfo {author} {\bibfnamefont {L.-H.}\ \bibnamefont {Mo}},\
  and\ \bibinfo {author} {\bibfnamefont {S.}~\bibnamefont {Ryu}},\ }\href
  {https://arxiv.org/abs/2507.03768} {\bibinfo {title} {{Entanglement
  transitions in structured and random nonunitary Gaussian circuits}}}
  (\bibinfo {year} {2025}{\natexlab{a}}),\ \Eprint
  {https://arxiv.org/abs/2507.03768} {arXiv:2507.03768} \BibitemShut {NoStop}%
\bibitem [{\citenamefont {Lapierre}\ \emph
  {et~al.}(2025{\natexlab{b}})\citenamefont {Lapierre}, \citenamefont
  {Pelliconi}, \citenamefont {Ryu},\ and\ \citenamefont
  {Sonner}}]{Lapierre2025}%
  \BibitemOpen
  \bibfield  {author} {\bibinfo {author} {\bibfnamefont {B.}~\bibnamefont
  {Lapierre}}, \bibinfo {author} {\bibfnamefont {P.}~\bibnamefont {Pelliconi}},
  \bibinfo {author} {\bibfnamefont {S.}~\bibnamefont {Ryu}},\ and\ \bibinfo
  {author} {\bibfnamefont {J.}~\bibnamefont {Sonner}},\ }\bibfield  {title}
  {\bibinfo {title} {Driven nonunitary dynamics of quantum critical systems},\
  }\href {https://doi.org/10.1103/lwrz-jxrr} {\bibfield  {journal} {\bibinfo
  {journal} {Phys. Rev. B}\ }\textbf {\bibinfo {volume} {112}},\ \bibinfo
  {pages} {104322} (\bibinfo {year} {2025}{\natexlab{b}})}\BibitemShut
  {NoStop}%
\bibitem [{\citenamefont {Bennett}\ \emph {et~al.}(2020)\citenamefont
  {Bennett}, \citenamefont {Melchers},\ and\ \citenamefont
  {Proppe}}]{Bennett2020}%
  \BibitemOpen
  \bibfield  {author} {\bibinfo {author} {\bibfnamefont {L.}~\bibnamefont
  {Bennett}}, \bibinfo {author} {\bibfnamefont {B.}~\bibnamefont {Melchers}},\
  and\ \bibinfo {author} {\bibfnamefont {B.}~\bibnamefont {Proppe}},\ }\href
  {http://dx.doi.org/10.17169/refubium-26754} {\bibinfo {title} {{Curta: A
  General-purpose High-Performance Computer at ZEDAT, Freie Universit{\"a}t
  Berlin}}} (\bibinfo {year} {2020})\BibitemShut {NoStop}%
\end{thebibliography}

%

\end{document}